\documentclass[pdflatex,sn-mathphys-ay]{sn-jnl}

\usepackage{graphicx}%
\usepackage{multirow}%
\usepackage{amsmath,amssymb,amsfonts}%
\usepackage{amsthm}%
\usepackage{mathrsfs}%
\usepackage{xcolor}
\usepackage{textcomp}%
\usepackage{manyfoot}%
\usepackage{booktabs}%
\usepackage{algorithm}%
\usepackage{algorithmicx}%
\usepackage{algpseudocode}%
\usepackage{listings}%

\usepackage{mathtools}%
\usepackage{nicefrac}%
\usepackage[utf8]{inputenc}%
\usepackage[T1]{fontenc}%
\usepackage{lmodern}%
\usepackage{array}%
\usepackage{tabularx}%
\usepackage{subcaption}%
\usepackage{tcolorbox}%
\usepackage{adjustbox}%
\usepackage{tikz}%
\usetikzlibrary{positioning,arrows,shapes}%
\usepackage{pgfplots}%
\pgfplotsset{compat=1.18}%

\graphicspath{{./Figures/}}%

\theoremstyle{thmstyleone}%
\newtheorem{theorem}{Theorem}[section]%

\theoremstyle{thmstyletwo}%
\newtheorem{example}[theorem]{Example}%

\theoremstyle{thmstylethree}%
\newtheorem{definition}[theorem]{Definition}%

\newcommand{\bx}{\mathbf{x}}%
\newcommand{\Pb}{\mathbb{P}}%
\newcommand{\Eb}{\mathbb{E}}%
\providecommand{\redit}[1]{#1}
\newcommand{\nedit}[1]{\textcolor{red}{#1}}
\begin{document}

\title[Controller-Augmented Hidden Markov Models]{Exact Constrained Decoding for Sequence Models: Controller-Augmented Hidden Markov Models and Beyond}

\author*[1]{\fnm{Lekha} \sur{Patel}}\email{lpatel@sandia.gov}
\author[2]{\fnm{Luis} \sur{Damiano}}\email{ladamia@sandia.gov}

\affil*[1]{\orgdiv{Scientific Machine Learning Department},
  \orgname{Sandia National Laboratories},
  \orgaddress{\city{Albuquerque}, \postcode{87123}, \state{NM},
  \country{USA}}}

\affil[2]{\orgdiv{AI Credibility    Department},
  \orgname{Sandia National Laboratories},
  \orgaddress{\city{Albuquerque}, \postcode{87123}, \state{NM},
  \country{USA}}}

\abstract{Hidden Markov models are foundational for sequential inference, yet conditioning on pathwise constraints such as precedence requirements, visitation cardinalities, or monotonic state progression can introduce long-range dependencies that require additional memory for dynamic programming. To deal with this, we present Controller-Augmented Hidden Markov Models (CHMMs), a framework that compiles each constraint into a finite-state controller tracking a sufficient finite history, after which standard forward--backward and Viterbi recursions on the augmented chain compute exact constrained posteriors and maximum a posteriori paths in discrete time and, through uniformization, maximum a posteriori augmented-state sequences at observation times in continuous time. While we establish several theoretical guarantees, a catalog of controller encodings for 11 constraint families operationalizes our framework. Empirically, we evaluate classical and neural decoders with and without controller augmentation on 3 real-world sequence-labeling tasks: gene-structure decoding in \emph{Drosophila melanogaster}, free-living activity recognition in CASAS smart homes, and protocol-defined activity recognition from wearable sensors. Together, theory and experiment characterize when exact controller augmentation is useful and when simpler approaches suffice.
}


\maketitle

\section{Introduction}
\label{sec:intro}
Hidden Markov Models constitute a foundational framework for probabilistic sequential modeling and inference. Their widespread deployment was first established in foundational applications including speech recognition \citep{rabiner1989tutorial}, computational biology \citep{durbin1998biological}, natural language processing \citep{manning1999foundations}, and computer vision \citep{ghahramani2001introduction}, and their flexibility and tractability have subsequently driven their continuous adoption in a vast array of modern domains. Recent systematic reviews \citep{Mor_2020, Dimri_2024} and a new generation of software \citep{Michelot_2025} highlight their vital role in fields as diverse as ecology \citep{Glennie_2022}, bioinformatics \citep{patel2019hidden,patel2021blinking}, cybersecurity \citep{Alshuaibi_2025}, human activity recognition \citep{Xue_2022}, financial risk assessment \citep{Surya_2024, Tan_2025}, industrial prognostics for fault diagnosis \citep{Li_2024}, and infrastructure monitoring \citep{Yamany_2025}. The enduring success of HMMs stems from their mathematical tractability, which enables efficient inference through dynamic programming whilst maintaining sufficient expressiveness to capture complex temporal dependencies \citep{baum1970maximization}. The Markovian assumption that underlies this tractability, however, becomes a fundamental limitation when sequential processes are governed by global pathwise constraints that create long-range dependencies incompatible with the memoryless property of the underlying chain.

The incorporation of domain-specific pathwise constraints into sequential inference represents a critical challenge across numerous application domains, and the constraints themselves manifest in a rich variety of forms. A first class involves temporal and duration requirements, ranging from operational cool-down periods in manufacturing \citep{lee2014manufacturing} and minimum observation periods in clinical care, to explicit duration modeling required in modern real-time systems \citep{hsmm2025recent}, financial forecasting \citep{assetmanagement2024semi, Tan_2025}, and human activity recognition \citep{hsmm2025activity, Xue_2022}. A second class involves precedence and ordering, which is foundational not only to clinical interventions but also to task-sequencing in robotics \citep{robotics2025optimal}, assembly planning \citep{salii2018precedence}, and logistics. A third, rapidly growing class involves safety, avoidance, and mutual exclusion, which is a central theme in modern autonomous systems and is typically formalized through Safe Reinforcement Learning and Constrained Markov Decision Processes \citep{cmdpsurvey2025, ijcaia2024safeRL, smith1999mutex}. A fourth class encodes complex logical or counting properties, such as the $k$-segment constraints used in signal analysis \citep{titsias2016statistical}, monotonicity required for interpretable machine learning \citep{xgb2025monotonicity}, and structural constraints in biological sequences \citep{yoon2009hidden}. These diverse applications share a fundamental requirement for exact probabilistic inference over trajectories that satisfy pathwise constraints transcending first-order Markovian dependencies, whilst differing crucially in whether the dominant constraint is purely local (a forbidden transition or a no-reentry rule) or genuinely cumulative (a precedence chain or an exactly-$k$-visit cardinality), a distinction that bears upon, without by itself determining, whether approximate decoding strategies suffice in practice.

This paper introduces Controller-Augmented Hidden Markov Models (CHMMs), a principled framework for exact inference under structured sequence labeling subject to finite-memory path constraints, namely those constraints that admit encoding by a deterministic finite controller. Our approach transforms constrained non-Markovian problems into unconstrained Markovian problems on an augmented state space through systematic controller design, with the controller maintaining a finite summary of the history that is sufficient to determine constraint satisfaction at each step of the sequential process. Subsequently, the framework recovers computational tractability whilst guaranteeing that all considered trajectories satisfy the specified constraints, and admits a clean treatment of both inference and parameter learning on the augmented state space.

\subsection{Main Contributions}
\label{sec:contributions-label}

Our work makes four contributions to the field of constrained sequential inference. First, we prove that the controller-augmentation construction preserves the Markov property whilst enforcing finite-memory pathwise constraints, enabling direct application of standard HMM algorithms on the augmented state space. Second, we provide a systematic methodology for controller design, with a catalog of eleven constraint families spanning the ordering, visitation, path, and temporal categories, each with an off-the-shelf encoding. Third, we present forward--backward, Viterbi, and constrained EM procedures on the augmented chain, prove their exactness and monotone-ascent properties, characterize their discrete-time complexity as linear in constraint size, and extend the algorithmic treatment to continuous time through uniformization. Fourth, we show empirically that the controller is separable from the model that supplies the sequence score, since in holding fitted neural scores fixed on three real-world tasks of substantively different character, we compare exact controller decoding against unconstrained decoding and against a bounded constrained beam. We find that exact decoding improves mean accuracy over both on the protocol-defined wearable-sensor task, is matched in accuracy by the beam on the remaining two, and on gene transcripts differs from that beam principally in that it never fails to return a feasible path, thereby furnishing an empirical account of when exact controller augmentation is necessary and when simpler approaches suffice.

\subsection{Related Work}
\label{sec:related}
Existing approaches to constrained sequential inference have pursued four distinct strategies, each with inherent limitations that motivate our work. For instance, post-hoc filtering methods apply constraints \emph{after} standard HMM inference, resulting in computational inefficiency since resources are expended on infeasible trajectories that must subsequently be discarded \citep{sridharan2010planning}. Such approaches fail to leverage constraint information \emph{during} inference, leading to suboptimal computational resource utilization.

Other approaches handle constraints through model modification, encoding constraints through manipulation of transition probabilities or extension to semi-Markov models \citep{yu2010hidden}. The hidden semi-Markov model (HSMM) framework is a standard method for modeling state duration constraints and remains a highly active field of research \citep{hsmm2025recent}. Whilst powerful for duration modeling, this entire class of models is not designed to handle general logical pathwise constraints, such as precedence relationships or complex counting requirements, that our framework addresses.

Approximate inference methods, including particle filters and beam searches, can incorporate constraints during sampling yet sacrifice the exactness guarantees that are crucial for applications requiring optimal solutions \citep{fearnhead2003line}. The beam-search-with-rejection strategy, in particular, performs beam decoding in which transitions to states violating locally-checkable constraints are pruned at each step \citep{deutsch2019general}, and has been deployed in natural language processing (NLP)-style structured prediction. However, whilst the strategy enforces admissibility against local constraints by construction, a beam of bounded width carries no guarantee for cumulative path requirements, since every retained prefix may foreclose the structure that the constraint demands, upon where the search either returns no feasible path or a feasible path that is not the highest scoring one. This category also includes modern Variational Inference (VI) techniques, which provide a flexible but often non-exact alternative \citep{vi2024modern}. All such methods trade the guarantee of optimality for computational tractability, a compromise our framework avoids for a broad class of constraints. 

A fourth strategy abandons the generative HMM framework altogether in favor of discriminative alternatives, of which the linear-chain conditional random field (CRF) is the canonical example \citep{lafferty2001conditional}. CRFs model the conditional distribution of label sequences given observations through feature functions over local cliques, supporting exact inference at fixed parameters and admitting the inclusion of hand-crafted feature templates that penalize locally-forbidden label transitions. Whilst this featurization renders CRFs effective for constraints expressible through adjacent-pair indicators, adjacent-pair features alone cannot represent cumulative path properties such as exactly-$k$-visit cardinalities or precedence chains on the original label space. This limitation is one of the label space and not of the model class, however, since a linear-chain CRF may be intersected with a finite-state acceptor to impose regular constraints in both training and decoding, as in the regular-constrained CRF of \cite{papay2022regular}, which is the closest antecedent of the construction studied here and which rests on the same product of a sequence model with an automaton that underlies weighted finite-state methods \citep{mohri2002weighted}. Our treatment differs in that we work with the generative likelihood, for which the feasibility event carries probability mass that depends on the parameters, and subsequently address the conditioned path measure, EM for the joint observation and feasibility likelihood, continuous-time jump constraints, and the sensitivity of the posterior to a misspecified constraint set.

Modern structured prediction increasingly relies on neural sequence models such as Long-Short Term Memory (LSTM)--CRF hybrids \citep{lample2016neural} and transformer-based taggers \citep{devlin2019bert}, which incorporate constraints as soft features through the loss function or via restricted softmax masking at inference time. Soft penalties and local masks of this kind do not guarantee that decoded trajectories satisfy global cumulative requirements. The present work targets the regime in which \emph{exact} enforcement is required, and because the augmented recursion consumes only per-position scores together with pairwise, initial and terminal terms, the controller composes directly with such scorers, which we exploit in Section~\ref{sec:experiments}. 

Constraints are also a central topic in parallel fields. In the area of formal methods, Probabilistic Model Checking verifies whether a Markov model satisfies a given temporal logic specification, such as those written in Linear Temporal Logic (LTL) \citep{pnueli1977temporal} or Probabilistic Computation Tree Logic \citep{kwiatkowska2022probabilistic}, and the systematic translation of temporal logic specifications into executable controllers is a mature field in its own right \citep{fainekos2006ltl}. This field, however, addresses the \emph{verification problem} (i.e. ``what is the probability that a trajectory satisfies a certain set of constraints?'') rather than the \emph{inference problem} (``given that trajectories must satisfy the same set of constraints, what is the most likely state sequence?''). In reinforcement learning, Constrained Markov Decision Processes are a major focus for safe planning \citep{cmdpsurvey2025}, but this work is concerned with finding an optimal policy rather than with performing inference on hidden data. More broadly, our work contributes to the field of structured inference, which uses problem-specific knowledge to create tractable models, as seen in highly-cited work on state-space models \citep{krishnan2017structured}.

Recent work has made progress on specific constraint families. \cite{titsias2016statistical} introduce $k$-segment constraints that restrict the number of state transitions, and this work is a key precedent for the state-augmentation technique we exploit in this paper, proving that an auxiliary counting chain preserves the joint distribution. Subsequently, this approach is limited to this specific constraint family and cannot express precedence constraints, mutual exclusions, or complex visitation requirements. \cite{christiansen2010inference} present a constraint logic programming framework (PRISM) for extending HMMs with side-constraints, whilst \cite{deutsch2019general} propose to express constraints as automata traversed during inference to guide search toward valid outputs. The latter approach is conceptually similar to what we define as a \emph{controller}, but it focuses on left-to-right sequential inference for NLP and lacks the complete theoretical framework for general probabilistic inference, such as proofs of EM convergence or complexity bounds. 

The product of a probabilistic sequence model with a finite automaton is therefore well established, and related ideas have appeared in forms specific to a constraint family or a model class, including auxiliary state augmentation for a single counting constraint~\citep{titsias2016statistical}, automata-guided decoding for NLP inference~\citep{deutsch2019general}, regular-language constraints on CRFs~\citep{papay2022regular}, and transition gating for precedence in robotic planning~\citep{robotics2025optimal}. What we contribute here is its systematic statistical treatment for generative hidden Markov models in discrete and continuous time: a formal encoding together with a catalog of controllers spanning the ordering, visitation, path, and temporal constraint families. Further, we provide an analysis of the augmented system's exactness, of the monotone ascent of EM for the joint observation and feasibility likelihood, of computational complexity, and of posterior sensitivity to constraint misspecification, with a controlled empirical comparison in which the decoding rule is isolated from the scoring model.

It is also important to differentiate our work from the established field of constrained parameter estimation. Most ``Constrained EM'' approaches operate either by projecting parameters back onto a valid set during the maximization step \citep{compstats2012projection, neurips2004projection}, or by injecting constraints on the latent-variable posteriors \citep{neurips2007posterior}. Our framework is fundamentally different. By augmenting the state space with a controller, we enforce path constraints directly on the model's structure, which subsequently enables the expectation step to perform exact inference over the valid trajectory space and ensures that the standard EM procedure \citep{dempster1977em} optimizes a likelihood that is, by construction, zero for all constraint-violating paths. The likelihood is the joint likelihood of the observations and the feasibility event, which is to be distinguished from the likelihood conditional on feasibility that arises when only feasible trajectories are sampled.


\subsection{Impact}
Our complexity analysis characterizes the computational cost of constrained inference as a function of the constraint specification, establishing that discrete-time inference is linear in the number of controller states at fixed base dimension, while making explicit that the controller itself may grow exponentially in the number of labels it must track, as in the all-different family. These results enable practitioners to assess the computational implications of a candidate constraint specification at the model-design stage, well before any inference or learning is undertaken.

The practical value of our approach is illustrated through synthetic experiments that examine exactness, computational scaling, and sensitivity to constraint misspecification. We additionally develop a unified evaluation framework comprising four headline metrics (position-wise accuracy, macro-averaged $F_{1}$, sequence-level trajectory validity rate, and segment-$F_{1}$) that operationalize structural fidelity alongside the standard classification metrics. The framework is subsequently exercised on three real-world sequence-labeling tasks.


\subsection{Paper Organization}
This paper is organized as follows. Section~\ref{sec:framework} establishes the mathematical framework for Controller-Augmented Hidden Markov Models, including formal problem formulation, the controller-augmentation construction, and the resulting Markovianization theorem. Section~\ref{sec:algorithms} presents algorithmic implementations for discrete-- and continuous--time systems, including modified forward--backward and Viterbi procedures together with the constrained EM algorithm and complexity analysis. Section~\ref{sec:theoretical} establishes the framework's robustness to constraint misspecification through a total-variation bound on the constrained posterior. Section~\ref{sec:metrics} introduces the unified evaluation framework, and Section~\ref{sec:experiments} subsequently presents the empirical evaluation across synthetic and real-world domains. Section~\ref{sec:discussion} concludes with a discussion on limitations, situating the work against alternative methodologies, and outlining future research directions. 
\section{Controller-Augmented Hidden Markov Model Framework}
\label{sec:framework}
This section establishes the mathematical framework for Controller-Augmented Hidden Markov Models (CHMMs). We begin by formalizing the constrained inference problem in both discrete and continuous time, subsequently introducing the controller-augmentation construction and proving that it preserves the Markov property whilst enforcing finite-memory pathwise constraints by construction. Within the framework, constraints assume a dual role in that they are not merely obstacles to inference, but rather a structured source of domain knowledge that may be exploited for parameter learning. Indeed, recent work has demonstrated that the incorporation of expert knowledge can prove critical for overcoming the data limitations and non-identifiability issues that frequently plague HMM estimation~\citep{ieee2024expert}, and the framework developed here provides a principled mechanism through which such knowledge may be injected directly into the model structure, thereby enforcing constraint satisfaction by construction rather than through approximation or post-hoc correction.

\subsection{Problem Formulation and Notation}
\label{sec:problem}
We consider a stochastic process $\{X_t : t \in \mathcal{T}\}$ taking values in a finite state space $\mathcal{X} = \{1, \ldots, S\}$ of cardinality $S$, where the time index set is $\mathcal{T} = \{0, \ldots, T\}$ for discrete-time processes and $\mathcal{T} = [0, T]$ for continuous-time processes. Associated with this hidden process are observations $\{Y_t : t \in \mathcal{T}_{\mathrm{obs}}\}$ taking values in an observation space~$\mathcal{Y}$, where $\mathcal{T}_{\mathrm{obs}} \subseteq \mathcal{T}$ denotes the set of observation times. Throughout the paper we adopt the standard convention of denoting random variables by capital letters and their realizations by lowercase letters, and we summarize all notation used in Table~\ref{tab:notation} of the Supplementary Materials. 

In a standard HMM~\citep{rabiner1989tutorial}, the hidden process satisfies the Markov property in the sense that the future evolution depends only on the current state, whilst the observation process $Y_t$ is conditionally independent given the hidden states, with emissions drawn from a parametric family $B = \{B_x : x \in \mathcal{X}\}$ for which $\Pb(y \mid X = x)$ has density (or mass function) $B_x(y)$. The dynamics are governed by transition kernels: in discrete time, the transition probability matrix $P = (P_{ij})_{i,j \in \mathcal{X}}$ has entries $P_{ij} = \Pb(X_{t+1} = j \mid X_t = i)$, whereas in continuous time the generator matrix $Q = (Q_{ij})_{i,j \in \mathcal{X}}$ has off-diagonal entries $Q_{ij} = \lim_{\delta \to 0} \Pb(X_{t+\delta} = j \mid X_t = i)/\delta$ for $i \neq j$ and diagonal entries $Q_{ii} = -\sum_{j \neq i} Q_{ij}$ that conserves total probability flow.

Numerous applications, however, require pathwise constraints that induce dependencies on the entire trajectory history and thereby violate the standard Markov property. We formalize such a constraint as a subset $\chi$ of feasible trajectories in the relevant path space. In discrete time, for example, $\chi \subseteq \mathcal{X}^{T+1}$. The constrained posterior distribution over trajectories then takes the form
\begin{equation}
\label{eq:constrained-posterior}
\adjustbox{max width=\columnwidth}{$\displaystyle
\Pb(X_{0:T} = \mathbf{x} \mid Y_{1:T} = \mathbf{y},\, X_{0:T}\in\chi)
\;\propto\;
p_{\mathrm{unc}}(\mathbf{x})\;\Pb(Y_{1:T} = \mathbf{y} \mid X_{0:T} = \mathbf{x})\;\mathbf{1}\{\mathbf{x} \in \chi\}
$}
\end{equation}
where $\mathbf{x} = (x_0, \ldots, x_T)$ and $\mathbf{y} = (y_1, \ldots, y_T)$, and where $p_{\mathrm{unc}}(\mathbf{x})$ denotes the unconstrained path measure, which in discrete time is given by
\begin{equation}
\label{eq:dtmc_path}
p_{\mathrm{unc}}(\mathbf{x}) = p_0(x_0)\prod_{t=0}^{T-1} P_{x_t, x_{t+1}}.
\end{equation}
In continuous time, a sample path on $[0,T]$ is specified by the number of jumps $N \ge 0$, jump times $0 = t_0 < t_1 < \cdots < t_N < T$, and visited states $x_0, \ldots, x_N \in \mathcal{X}$ with $X_t = x_n$ for $t \in [t_n, t_{n+1})$ and $t_{N+1} := T$, and the unnormalized path density of the continuous-time Markov chain (CTMC) with generator $Q$ becomes
\begin{equation}
\label{eq:ctmc_path_density}
\adjustbox{max width=\columnwidth}{$\displaystyle
p_{\mathrm{unc}}(\mathbf{x}) = p_0(x_0)\left(\prod_{n=1}^{N} Q_{x_{n-1},x_n}\right)
\exp\!\left(\sum_{n=0}^{N} Q_{x_n,x_n}(t_{n+1} - t_n)\right)
$}
\end{equation}

The objective of this paper is to develop a unified computational framework that enforces finite-memory pathwise constraints whilst preserving the efficiency of Markovian inference algorithms. Our approach rests on the principle that non-Markovian processes with long-range dependencies can be transformed into first-order Markovian processes on a suitably enlarged state space~\citep{jpo2003longmemory}, and we realize this systematically by augmenting the base state with a deterministic \emph{controller}, a technique that draws on ideas from state-space abstraction~\citep{jair2018abstractions} and from auxiliary counting variables previously introduced for specific constraints~\citep{titsias2016statistical}.

\subsection{Controller Augmentation and Markovianization}
\label{sec:controller}
To restore the Markov property under pathwise constraints, we augment the state space with a \emph{controller} variable that tracks a sufficient finite summary of the history for constraint verification. This approach is formally grounded in supervisory control theory, which defines a supervisor (in our setting, the controller) to ensure that a discrete-event process (in our setting, the HMM) adheres to a specified target language (in our setting, the constraint set)~\citep{wonham1987sct}.

\begin{definition}[Controller specification and killed augmented process]
\label{def:controller}
A \emph{controller specification} is a tuple $(\mathcal{C}, c_0, \tau, F, \mathcal{F}_T)$ comprising:
\begin{enumerate}
  \item \textbf{Controller state space:} a finite set $\mathcal{C}$;
  \item \textbf{Initialisation:} a function $c_0\colon\mathcal{X}\to\mathcal{C}$ such that $C_0 = c_0(X_0)$;
  \item \textbf{Update:} a deterministic transition rule $C_{t+1} = \tau(C_t, X_t, X_{t+1}, t)$ in discrete time, or the analogous rule applied at jump times in continuous time;
  \item \textbf{Gating:} a feasibility function $F : \mathcal{C}\times \mathcal{X}\times \mathcal{X}\times \mathcal{T} \to \{0,1\}$ which specifies, for each augmented transition $(i,c)\to(j,c')$ at time $t$, that the transition is locally admissible if and only if $F(c,i,j,t)=1$ and $c'=\tau(c,i,j,t)$;
  \item \textbf{Terminal acceptance:} a subset $\mathcal{F}_T \subseteq \mathcal{C}$ specifying the controller states at which the trajectory is deemed cumulatively feasible.
\end{enumerate}
For time-homogeneous constraints we suppress the explicit dependence on $t$ and write $\tau(c,i,j)$ and $F(c,i,j)$. The \emph{live augmented state} is $\tilde{X}_t = (X_t, C_t) \in \tilde{\mathcal{X}} = \mathcal{X} \times \mathcal{C}$, and the \emph{killed process} $\bar{X}_t \in \bar{\mathcal{X}} = \tilde{\mathcal{X}} \cup \{\bot\}$ extends the live state space with an absorbing dead state $\bot$ to which any infeasible transition sends the process. The associated \emph{feasibility event} is
\[
\mathsf{Acc} = \{\bar{X}_t \neq \bot\;\forall\, t \le T\} \cap \{C_T \in \mathcal{F}_T\}.
\]
\end{definition}

The constraint set $\chi$ does not appear directly in Definition~\ref{def:controller}, rather, the controller specification is connected to the constraint of interest through an encoding design, defined in Definition \ref{def:encoding}.

\begin{definition}[Encoding of a constraint]
\label{def:encoding}
A controller specification $(\mathcal{C}, c_0, \tau, F, \mathcal{F}_T)$ is said to \emph{encode} the constraint set $\chi \subseteq \mathcal{X}^{\mathcal{T}}$ if, for every trajectory $\mathbf{x} = x_{0:T}$ with associated controller trace defined by $c_0 = c_0(x_0)$ and $c_{t+1} = \tau(c_t, x_t, x_{t+1}, t)$, the equivalence
\[
\mathbf{x} \in \chi
\;\;\Longleftrightarrow\;\;
\bigl(F(c_t, x_t, x_{t+1}, t) = 1, \; 0 \le t < T\bigr) \;\wedge\; \bigl(c_T \in \mathcal{F}_T\bigr)
\]
holds. In what follows, we assume that the controller is chosen so as to encode the constraint of interest in this manner.
\end{definition}

In this paper, all inference and learning are performed under the conditional distribution given $\mathsf{Acc}$, denoted $p(\cdot \mid \mathsf{Acc})$, where the constrained path law equivalently satisfies $p(\mathbf{x} \mid \mathsf{Acc}) \propto p(\mathbf{x})\,\mathbf{1}\{\mathbf{x} \in \chi\}$. The controller therefore enforces the constraint through two complementary mechanisms: \emph{local gating} via $F(c,i,j)$ blocks transitions that would violate path-dependent requirements given the current controller state, whilst \emph{terminal acceptance} via $\mathcal{F}_T$ restricts the inference normalization to trajectories whose terminal controller state lies in $\mathcal{F}_T$, thereby ensuring that cumulative requirements are satisfied. We further assume throughout that the controller state $c$ affects the dynamics but not the observation distribution, so as to preserve the interpretability of the base state $x$ at observation time.

We now formally establish the augmented kernels and prove that the construction preserves the Markov property whilst enforcing constraints exactly.

\begin{theorem}[Killed augmented kernels: discrete and continuous time]
\label{thm:augmented_kernels}
Let $X$ be a base Markov chain with transition matrix $P$ in discrete time, or generator $Q$ in continuous time, and let $(\mathcal{C}, c_0, \tau, F, \mathcal{F}_T)$ be a controller specification that encodes the constraint set $\chi$ in the sense of Definition~\ref{def:encoding}.

\medskip\noindent
\textbf{Discrete time.} The \emph{live} augmented kernel on $\tilde{\mathcal{X}}$ defined by
\begin{equation}
\label{eq:live_kernel}
\tilde{P}_{(i,c),(j,c')}
  \;=\; P_{ij}\;\mathbf{1}\{F(c,i,j)=1\}\;\mathbf{1}\{c'=\tau(c,i,j)\}
\end{equation}
is in general sub-stochastic. The corresponding killed kernel on $\bar{\mathcal{X}} = \tilde{\mathcal{X}} \cup \{\bot\}$ is given by
\begin{align}
\bar{P}_{(i,c),(j,c')} &= \tilde{P}_{(i,c),(j,c')}, \label{eq:barP_live}\\
\bar{P}_{(i,c),\bot}    &= 1 - \sum_{(j,c') \in \tilde{\mathcal{X}}} \tilde{P}_{(i,c),(j,c')}, \label{eq:barP_kill}\\
\bar{P}_{\bot,\bot}      &= 1. \label{eq:barP_cem}
\end{align}
Then $(\bar{X}_t)$ is a Markov chain on $\bar{\mathcal{X}}$, and the constrained posterior satisfies
\[
p(\mathbf{x} \mid \mathbf{y}, \mathsf{Acc})
  \;\propto\; p(\mathbf{x})\,p(\mathbf{y}\mid \mathbf{x})\,\mathbf{1}\{\mathbf{x}\in\chi\}.
\]

\medskip\noindent
\textbf{Continuous time.} The killed generator $\bar{Q}$ on $\bar{\mathcal{X}}$ defined by
\begin{align}
\bar{Q}_{(i,c),(j,c')}
  &= Q_{ij}\;\mathbf{1}\{F(c,i,j)=1\}\;\mathbf{1}\{c'=\tau(c,i,j)\},
  \quad j \neq i,
  \label{eq:barQ_live}\\
\bar{Q}_{(i,c),\bot}
  &= \sum_{j \neq i} Q_{ij}\;\mathbf{1}\{F(c,i,j)=0\},
  \label{eq:barQ_kill}\\
\bar{Q}_{(i,c),(i,c)}
  &= -\sum_{\substack{(j,c') \in \tilde{\mathcal{X}} \\ (j,c')\neq(i,c)}} \bar{Q}_{(i,c),(j,c')} \;-\; \bar{Q}_{(i,c),\bot},
  \label{eq:barQ_diag}\\
\bar{Q}_{\bot,\bot} &= 0,
  \qquad \bar{Q}_{\bot,z} = 0 \;\text{ for } z \neq \bot
  \label{eq:barQ_cem}
\end{align}
is a valid generator, and the same  characterization of the constrained posterior holds.
\end{theorem}


For continuous-time inference, we uniformize the killed generator $\bar{Q}$ at any rate $\lambda \ge \max_{i \in \mathcal{X}}(-Q_{ii})$, obtaining the uniformized transition matrix $\bar{P}_{\mathrm{unif}} = I + \bar{Q}/\lambda$, with the transition matrix over an interval of length $\Delta t$ given by $\bar{P}(\Delta t) = \exp(\bar{Q}\,\Delta t)$.

\subsection{Constraint Catalog}
\label{sec:catalog}
Having established the controller-augmentation construction, we now present a catalog of common pathwise constraints together with their systematic controller encodings. These constraints are well-established in formal methods, scheduling, and temporal planning, where concepts such as precedence~\citep{salii2018precedence,smith1999mutex} and temporal logic~\citep{pnueli1977temporal} have been extensively formalized, and our framework compiles these logical specifications into a probabilistic inference-ready format. Table~\ref{tab:constraint-catalog-complete} provides the catalog, in which each constraint is specified by its controller space $\mathcal{C}$, update function $\tau$, gating condition $F$, and terminal acceptance set $\mathcal{F}_T$. The notation $\mathcal{F}_T = \mathcal{C}$ indicates that all controller states are valid at termination, whereas specific subsets such as $\mathcal{F}_T = \{K\}$ enforce cumulative requirements.

\begin{table*}[t!]
\centering
\caption{Catalogue of pathwise constraints and their controller encodings under the framework of Definition~\ref{def:controller}. Each row specifies the controller state space $\mathcal{C}$, its size, the update function $\tau(c,i,j)$, the local-feasibility gate, and the terminal acceptance set $\mathcal{F}_T$, with the initialization $c_0$ omitted from the table for compactness and given in the worked examples or in the Supplementary Materials where relevant. The full dense-transition complexity of constrained inference under any of these encodings is $O(TS^{2}|\mathcal{C}|)$, whilst the sparse-transition complexity is $O(TS|\mathcal{C}|d_{\mathrm{avg}})$, where $d_{\mathrm{avg}}$ denotes the average out-degree of the augmented transition graph.}
\label{tab:constraint-catalog-complete}
\resizebox{\textwidth}{!}{
\begin{tabular}{@{}llcllll@{}}
\toprule
\textbf{Constraint} & \textbf{Controller} $\mathcal{C}$ & \textbf{Size} & \textbf{Update} $\tau(c,i,j)$ & \textbf{Gate} $F(c,i,j) = 0$ when & \textbf{Terminal} $\mathcal{F}_T$ \\
\midrule
\multicolumn{6}{l}{\textit{Ordering Constraints}} \\
Precedence $a \prec b$ & $\{0,1\}$ (visited $a$?) & $2$ & $c \vee \mathbf{1}\{j = a\}$ & $j = b \wedge c = 0$ & $\mathcal{C}$ \\
Forbidden edges $\mathcal{B}$ & None (memoryless) & $1$ & --- & $(i,j) \in \mathcal{B}$ & $\mathcal{C}$ \\
Stage monotone & $\{0,\ldots,G{-}1\}$ & $G$ & $\max\{c, k(j)\}$ & $k(j) > c + 1$ & $\mathcal{C}$ \\
\midrule
\multicolumn{6}{l}{\textit{Visitation Constraints for $A \subseteq \mathcal{X}$}} \\
At-least-$K$ visits to $A$\,$^{\dagger}$ & $\{0,\ldots,K\}$ & $K{+}1$ & $\min\{K,\, c + \varepsilon_A(i,j)\}$ & None & $\{K\}$ \\
At-most-$K$ visits to $A$ & $\{0,\ldots,K\}$ & $K{+}1$ & $\min\{K,\, c + \varepsilon_A(i,j)\}$ & $\varepsilon_A(i,j) = 1 \wedge c = K$ & $\mathcal{C}$ \\
Exactly-$K$ visits to $A$ & $\{0,\ldots,K\}$ & $K{+}1$ & $\min\{K,\, c + \varepsilon_A(i,j)\}$ & $\varepsilon_A(i,j) = 1 \wedge c = K$ & $\{K\}$ \\
\midrule
\multicolumn{6}{l}{\textit{Path Constraints}} \\
$K$-transition (so $K{+}1$ segments) & $\{0,\ldots,K\}$ (transitions) & $K{+}1$ & $c + \mathbf{1}\{i \neq j\}$ & $i \neq j \wedge c = K$ & $\{K\}$ \\
All-different & $\mathcal{P}(\mathcal{X})$ (visited set) & $2^S$ & $c \cup \{j\}$ & $j \in c$ & $\mathcal{C}$ \\
\midrule
\multicolumn{6}{l}{\textit{Temporal Constraints}} \\
No-dwell in $A$ & None (memoryless) & $1$ & --- & $i \in A \wedge j \in A$ & $\mathcal{C}$ \\
No-reentry to $A$\,$^{\ddagger}$ & $\{0,1,2\}$ (phase) & $3$ & see footnote $\ddagger$ & $c = 2 \wedge j \in A$ & $\mathcal{C}$ \\
Cool-down $\Delta$ on $A$\,$^{\S}$ & $\{0,\ldots,\Delta\}$ (timer) & $\Delta{+}1$ & see footnote $\S$ & $\varepsilon_A(i,j) = 1 \wedge c > 0$ & $\mathcal{C}$ \\
\bottomrule
\end{tabular}
}
\vspace{0.4em}
\\
\begin{minipage}{0.97\textwidth}
\footnotesize
$^{\dagger}$ Because the update saturates at $K$, the controller never exceeds $K$, so the terminal set $\{K\}$ is operationally equivalent to $\{c \in \mathcal{C} : c \ge K\}$.
\\
$^{\ddagger}$ The no-reentry update tracks three phases: $0 = $ never in $A$, $1 = $ currently in $A$, $2 = $ exited $A$. Explicitly, $\tau(c,i,j) = 1$ if $j \in A$ and $c \in \{0,1\}$; $\tau(c,i,j) = 2$ if $c = 1$ and $j \notin A$; $\tau(c,i,j) = c$ otherwise. State $2$ is absorbing.
\\
$^{\S}$ The cool-down update sets $\tau(c,i,j) = \Delta$ when $i \in A$ and $j \notin A$ (cool-down begins on exit from $A$), and $\tau(c,i,j) = \max\{0,\, c - 1\}$ otherwise (timer decrements each step until zero).
\end{minipage}
\end{table*}
The catalogue makes clear that diverse constraint families admit systematic encodings of modest controller complexity. Memoryless constraints such as forbidden edges and no-dwell require no additional state ($|\mathcal{C}| = 1$) and therefore preserve the original inference cost, whilst constraints with finite memory scale linearly with their natural parameters. For example, visitation counts require $O(K)$ controller states and temporal constraints require $O(\Delta)$ states for timing windows of length $\Delta$. This linear scaling ensures computational tractability even for complex constraint combinations, and the systematic nature of the encoding further enables practitioners to compose multiple constraints through controller products, with overall complexity scaling as the product of the individual controller sizes.

To illustrate how these encodings are realized in practice, and to exhibit the two enforcement mechanisms in their canonical forms, we now present two worked examples: a precedence constraint whose enforcement is purely local (gating only), and a visitation constraint whose enforcement is purely cumulative (terminal acceptance only). Three further worked examples for the visitation, $k$-segment and cool-down constraints are deferred to Supplementary Materials Section~\ref{sec:catalog-examples-deferred}.

\subsubsection{Precedence Constraint}
\label{sec:example_precedence}
Precedence constraints are foundational in robotics for assembly planning~\citep{robotics2025optimal} and in logistics for task sequencing, where precedence is typically defined as a partial order over tasks~\citep{salii2018precedence}. We illustrate the encoding through a three-state system in which state~$3$ may be entered only after state~$1$ has been visited at least once.

\begin{example}[Precedence Constraint]
\label{ex:precedence}
Consider a base HMM with states $\mathcal{X} = \{1, 2, 3\}$ and transition matrix
\begin{equation}
\label{eq:P_prec}
P = \begin{pmatrix}
0.5 & 0.3 & 0.2 \\
0.2 & 0.6 & 0.2 \\
0.1 & 0.3 & 0.6
\end{pmatrix}.
\end{equation}
The constraint $\chi = \{x_{0:T} : \text{state 3 is entered only after state 1 has been visited}\}$ is encoded with controller space $\mathcal{C} = \{0, 1\}$ tracking whether state~$1$ has been visited, initialisation $c_0(x) = \mathbf{1}\{x = 1\}$, update $\tau(c, i, j) = c \vee \mathbf{1}\{j = 1\}$, gate $F(c, i, j) = \mathbf{1}(j \neq 3) \lor \mathbf{1}(c = 1)$, and terminal acceptance $\mathcal{F}_T = \mathcal{C}$.
\end{example}

The controller is a binary flag whose state at time $t$ records whether state~$1$ has been visited by then. Subsequently, the gate blocks any transition into state~$3$ whilst the flag is zero, and admits all such transitions once the flag has been set. The augmented transition kernel
\[
\tilde{P}_{(i,c),(j,c')}
  \;=\; P_{ij}\;\mathbf{1}\{F(c,i,j) = 1\}\;\mathbf{1}\{c' = \tau(c,i,j)\}
\]
therefore assigns probability zero to all transitions into state~$3$ until the flag transitions to one, after which the dynamics coincide with those of the base chain. Figure~\ref{fig:example-graph} shows the resulting augmented state graph, where the augmented kernel possesses a block-triangular structure
\begin{equation}
\label{eq:block_prec}
\tilde{P} = \begin{pmatrix}
\tilde{P}_{00} & \tilde{P}_{01} \\
\mathbf{0} & \tilde{P}_{11}
\end{pmatrix},
\end{equation}
in which $\tilde{P}_{00}$ governs transitions before state~$1$ is visited (with its third column zeroed out), $\tilde{P}_{01}$ captures transitions that visit state~$1$ for the first time, and $\tilde{P}_{11}$ equals the unconstrained matrix~$P$ since the precedence requirement is no longer active once the flag has been set.

\begin{figure}[h]
\centering
\begin{tikzpicture}[
    augstate/.style={rectangle, draw, rounded corners, thick, minimum width=1.5cm, minimum height=1cm},
    blockedstate/.style={rectangle, draw, rounded corners, thick, minimum width=1.5cm, minimum height=1cm, fill=red!20},
    goodarrow/.style={->, >=stealth, thick},
    badarrow/.style={->, >=stealth, thick, red, dashed},
    scale=0.8
]

\node[augstate] (s10) at (0, 3) {$(1,0)$};
\node[augstate] (s20) at (4, 3) {$(2,0)$};
\node[blockedstate] (s30) at (8, 3) {$(3,0)$};

\node[augstate] (s11) at (0, 0) {$(1,1)$};
\node[augstate] (s21) at (4, 0) {$(2,1)$};
\node[augstate] (s31) at (8, 0) {$(3,1)$};

\node[gray, above=1.1cm of s10] {$c=0$: not visited 1};
\node[gray, below=1.1cm of s11] {$c=1$: visited 1};
\node[red] at (6, 5.3) {$\times$ blocked};

\path[goodarrow] (s10) edge[loop above, looseness=10] node {0.5} (s10);
\path[goodarrow] (s10) edge[bend left=20] node[above] {0.3} (s20);
\path[badarrow] (s10) edge[bend left=50] node[above, red] {} (s30);

\path[goodarrow] (s20) edge[loop above, looseness=10] node {0.6} (s20);
\path[goodarrow] (s20) edge[bend right=30] node[left] {0.2} (s11);
\path[badarrow] (s20) edge[bend left=20] node[above, red] {} (s30);

\path[goodarrow] (s11) edge[loop below, looseness=10] node {0.5} (s11);
\path[goodarrow] (s11) edge[bend left=20] node[below] {0.3} (s21);
\path[goodarrow] (s11) edge[bend left=25] node[below] {0.2} (s31);

\path[goodarrow] (s21) edge[bend left=20] node[above] {0.2} (s11);
\path[goodarrow] (s21) edge[loop below, looseness=10] node {0.6} (s21);
\path[goodarrow] (s21) edge[bend left=20] node[below] {0.2} (s31);

\path[goodarrow] (s31) edge[bend left=50] node[below] {0.1} (s11);
\path[goodarrow] (s31) edge[bend left=20] node[above] {0.3} (s21);
\path[goodarrow] (s31) edge[loop below, looseness=10] node {0.6} (s31);
\end{tikzpicture}
\caption{Augmented state graph for the precedence constraint $1 \prec 3$. State $(3,0)$ is unreachable (shaded red), and transitions into state~$3$ from the upper row are blocked (dashed red) whilst the flag $c$ is zero. Once state~$1$ has been visited, the controller transitions to $c = 1$, and all transitions into state~$3$ become admissible.}
\label{fig:example-graph}
\end{figure}

\section{Algorithmic Framework}
\label{sec:algorithms}
Having established that the controller-augmentation construction transforms a constrained non-Markovian process into an unconstrained Markovian one on the augmented state space $\tilde{\mathcal{X}} = \mathcal{X} \times \mathcal{C}$, we now consider inference and learning. The central observation that organizes this section is that, because the killed augmented process is itself Markovian,  standard dynamic-programming machinery for hidden Markov models applies to the augmented kernel without modification, and the constrained posterior of interest may therefore be obtained by running ordinary forward--backward and Viterbi recursions on $\tilde{\mathcal{X}}$ and subsequently marginalizing the controller dimension. We develop the discrete-time recursions in Section~\ref{sec:inference}, indicate the corresponding continuous-time treatment in the same subsection by appeal to the uniformization construction of Theorem~\ref{thm:augmented_kernels}, and subsequently establish parameter learning through expectation-maximization in Section~\ref{sec:em_learning} and the associated complexity analysis in Section~\ref{sec:complexity}.
            
\subsection{Constrained Inference}
\label{sec:inference}
The constrained smoothing problem requires the posterior marginal $\Pb(X_t = i \mid Y_{1:T}, \mathsf{Acc})$, whilst the constrained MAP problem requires the most probable feasible trajectory $\arg\max_{\mathbf{x} \in \chi} \Pb(\mathbf{x} \mid Y_{1:T})$. Both quantities are obtained by adapting the standard forward--backward and Viterbi recursions, the only essential change being that the recursions operate on the augmented kernel $\tilde{P}$ of Equation~\eqref{eq:live_kernel} rather than on the base kernel~$P$.

\subsubsection{Forward--Backward Recursion}
\label{sec:forward_backward}
Define the constrained forward variable
\begin{equation}
\alpha_t(i, c) \; = \; \Pb(Y_{1:t},\, \bar{X}_t = (i, c)),
\label{eq:alpha_def}
\end{equation}
in which the prefix-feasibility event encompasses the requirement that no infeasible transition be taken on the first $t-1$ steps and is enforced through the gating function $F$ embedded in $\tilde{P}$. Because the controller initialization $C_0 = c_0(X_0)$ is deterministic, the initialization of the forward recursion is
\begin{equation}
\alpha_0(i, c) \;=\; \nu(i)\, \mathbf{1}\{c = c_0(i)\},
\label{eq:alpha_init}
\end{equation}
and the recursion for $t = 0, \dots, T-1$ takes the standard form on the augmented state space
\begin{equation}
\alpha_{t+1}(j, c') \;=\; B_j(y_{t+1}) \sum_{(i,c) \in \tilde{\mathcal{X}}} \alpha_t(i, c)\, \tilde{P}_{(i,c),(j,c')}.
\label{eq:alpha_rec}
\end{equation}
The constrained marginal likelihood is recovered by restricting the terminal sum to the accepting set
\begin{equation}
Z \;=\; \Pb(Y_{1:T},\, \mathsf{Acc}) \;=\; \sum_{i \in \mathcal{X}} \sum_{c \in \mathcal{F}_T} \alpha_T(i, c),
\label{eq:Z_def}
\end{equation}
which serves as the global normalizing constant for all subsequent posterior quantities. We emphasize that the rows of $\tilde{P}$ are not normalized: $\tilde{P}$ is sub-stochastic and the missing mass corresponds to killed (infeasible) trajectories, so renormalization would corrupt the constrained posterior by reweighting feasible trajectories to compensate for the absence of infeasible ones.

The backward variable is defined by
\begin{equation}
\beta_t(i, c) \;=\; \Pb(Y_{t+1:T},\, \bar{X}_T \neq \bot,\, C_T \in \mathcal{F}_T \mid \bar{X}_t = (i, c)),
\label{eq:beta_def}
\end{equation}
with terminal initialization
\begin{equation}
\beta_T(i, c) \;=\; \mathbf{1}\{c \in \mathcal{F}_T\}
\label{eq:beta_init}
\end{equation}
and recursion for $t = T-1, \ldots, 0$,
\begin{equation}
\beta_t(i, c) \;=\; \sum_{(j,c') \in \tilde{\mathcal{X}}} \tilde{P}_{(i,c),(j,c')}\, B_j(y_{t+1})\, \beta_{t+1}(j, c').
\label{eq:beta_rec}
\end{equation}
The constrained smoothed marginal on the augmented state is then
\begin{equation}
\gamma_t(i, c) \;=\; \frac{\alpha_t(i, c)\, \beta_t(i, c)}{Z}
\label{eq:gamma}
\end{equation}
and the corresponding marginal on the base state is obtained by summing the controller dimension out
\begin{equation}
\Pb(X_t = i \mid Y_{1:T}, \mathsf{Acc}) \;=\; \sum_{c \in \mathcal{C}} \gamma_t(i, c).
\label{eq:gamma_base}
\end{equation}
Equations~\eqref{eq:alpha_init}--\eqref{eq:gamma_base} together specify the constrained forward--backward algorithm, whereby pseudocode is provided as Algorithm~\ref{alg:constrained_forward_backward} in the Supplementary Materials for completeness. The key observation is that both passes have precisely the same per-step structure as ordinary forward--backward, with the augmented kernel $\tilde{P}$ replacing the base kernel and the terminal acceptance restriction $c \in \mathcal{F}_T$ replacing the unconstrained terminal condition $\beta_T \equiv 1$.

\subsubsection{Viterbi Recursion}
\label{sec:viterbi}
The constrained MAP trajectory is obtained by replacing the sum in the forward recursion with a maximum, in the standard manner. Define the Viterbi variable
\begin{equation}
\adjustbox{max width=\columnwidth}{$\displaystyle
\delta_t(i, c) \;=\; \max_{x_{0:t-1} \in \mathcal{X}^{t}}\; \Pb\bigl(Y_{1:t},\, \bar{X}_{0:t-1} = (x_{0:t-1}, c_{0:t-1}),\, \bar{X}_t = (i, c)\bigr),
$}
\label{eq:delta_def}
\end{equation}
in which the controller trace $c_{0:t-1}$ is determined deterministically from $x_{0:t-1}$ by $c_0 = c_0(x_0)$ and $c_{s+1} = \tau(c_s, x_s, x_{s+1})$ for $s \leq t-2$. The initialization is $\delta_0(i, c) = \nu(i)\, \mathbf{1}\{c = c_0(i)\}$, and the recursion takes the form
\begin{equation}
\delta_{t+1}(j, c') \;=\; B_j(y_{t+1}) \max_{(i, c) \in \tilde{\mathcal{X}}} \bigl[\delta_t(i, c)\, \tilde{P}_{(i,c),(j,c')}\bigr],
\label{eq:delta_rec}
\end{equation}
maintaining backpointers $\psi_{t+1}(j, c') = \arg\max_{(i,c)} \bigl[\delta_t(i, c)\, \tilde{P}_{(i,c),(j,c')}\bigr]$ to permit reconstruction. Subsequently, the optimal terminal augmented state is $\bigl(x_T^{\star}, c_T^{\star}\bigr) = \arg\max_{(j, c) : c \in \mathcal{F}_T} \delta_T(j, c)$, in which the maximization over the final controller index is restricted to $\mathcal{F}_T$ in order to enforce the cumulative component of the constraint, and the optimal trajectory is then recovered by backtracking $(x_t^{\star}, c_t^{\star}) = \psi_{t+1}(x_{t+1}^{\star}, c_{t+1}^{\star})$ for $t < T$. We provide the corresponding pseudocode in Algorithm~\ref{alg:constrained_viterbi} of the Supplementary Materials, working in log-space for numerical stability.

\subsubsection{Continuous-Time Treatment}
\label{sec:cthmm_inference}
The recursions of the previous two subsections transfer to the continuous-time setting directly through the uniformization construction of Theorem~\ref{thm:augmented_kernels}. Here, forward--backward and Viterbi proceed identically on $\tilde{\mathcal{X}}$, with the discrete-time kernel $\tilde{P}$ replaced by $\bar{P}(\Delta t_v) = \exp(\bar{Q}\,\Delta t_v)$ at each inter-observation interval.

\textbf{Forward--backward recursions.}
With observations arriving at irregular times $0 = t_0 < t_1 < \cdots < t_V = T$, the discrete-time forward recursion of Equation~\eqref{eq:alpha_rec} generalizes to
\begin{equation}
\alpha_{v+1}(j, c') \;=\; B_j(y_{t_{v+1}}) \sum_{(i, c) \in \tilde{\mathcal{X}}} \alpha_v(i, c)\, \bar{P}_{(i,c),(j,c')}(t_{v+1} - t_v),
\label{eq:alpha_ct}
\end{equation}
and the analogous backward recursion follows the same pattern. Restriction to the live state space $\tilde{\mathcal{X}}$ throughout the recursions is justified by the absorbing nature of $\bot$ in the killed generator, which guarantees that any mass leaving $\tilde{\mathcal{X}}$ does not return.

\textbf{Maximum a posteriori decoding.}
The continuous-time MAP question requires a precise formulation of the path object being optimized. We adopt the standard CT-HMM convention of decoding the most likely augmented-state sequence at observation times, namely
\begin{equation}
\label{eq:ctmap}
\hat{\bar{x}}_{0:V} \;\in\; \arg\max_{\bar{x}_{0:V}}\, \Pb\bigl(\bar{X}_{t_{0:V}} = \bar{x}_{0:V} \,\bigl|\, Y_{0:V},\, \mathcal{T}_{\mathrm{obs}},\, \mathsf{Acc}\bigr),
\end{equation}
in which the maximization is over augmented-state assignments at the observation times, with the latent path between observations integrated out under the augmented kernel $\bar{P}$. This is the natural continuous-time analogue of the discrete-time Viterbi object, and it admits a Viterbi-style recursion identical in structure to the discrete-time case with $\bar{P}(t_{v+1} - t_v)$ replacing the discrete-time transition matrix:
\begin{equation}
\adjustbox{max width=\columnwidth}{$\displaystyle
\delta_{v+1}(j, c') \;=\; B_{j}(y_{t_{v+1}}) \,\max_{(i, c) \in \tilde{\mathcal{X}}}\,\delta_{v}(i, c)\, \bar{P}_{(i, c),\, (j, c')}(t_{v+1} - t_v),
$}
\label{eq:delta_ct}
\end{equation}
with the optimal augmented-state assignment recovered by standard backtracking. The recursion runs in time $O(V\,|\tilde{\mathcal{X}}|^{2})$ per matrix-exponential evaluation, with the matrix exponentials themselves computed once per distinct inter-observation interval and cached. Decoding the full continuous-time jump path between observations is a separate question that can be addressed through the ``uniformized-virtual-jump'' representation~\citep{liu2015efficient}, but is not required for the constraint validity and segment-level metrics of Section~\ref{sec:metrics}, which are determined entirely by the augmented states at observation times.

\subsubsection{Sparsity in the Augmented Kernel}
\label{sec:sparsity}
The augmented kernel $\tilde{P}$ is in practice highly sparse, since the gating function~$F$ blocks all infeasible transitions and the controller update~$\tau$ is deterministic. Concretely, for each augmented source state $(i, c)$ at most one augmented destination state $(j, c')$ is reached for each base destination $j$, and the proportion of $j$ for which $F(c, i, j) = 1$ is typically small. We exploit this structure by precomputing, for each augmented source, the adjacency list
\begin{equation}
\mathcal{N}(i, c) \;=\; \bigl\{(j, c') \in \tilde{\mathcal{X}} : \tilde{P}_{(i,c),(j,c')} > 0\bigr\}
\label{eq:adjlist}
\end{equation}
and rewriting the inner loops of Algorithms~\ref{alg:constrained_forward_backward} and~\ref{alg:constrained_viterbi} to iterate over $\mathcal{N}(i, c)$ rather than over $\mathcal{X} \times \mathcal{C}$. The resulting per-step cost is $O\bigl(S\, |\mathcal{C}|\, d_{\mathrm{avg}}\bigr)$, where $d_{\mathrm{avg}} = (S\,|\mathcal{C}|)^{-1} \sum_{(i,c)} |\mathcal{N}(i, c)|$ is the mean out-degree of the augmented transition graph, against the dense-matrix cost of $O\bigl(S^{2}\, |\mathcal{C}|^{2}\bigr)$, with the saving substantial whenever $d_{\mathrm{avg}} \ll S\, |\mathcal{C}|$.

\subsection{Parameter Learning via Expectation-Maximization}
\label{sec:em_learning}
The parameters of a CHMM may be learned from data using a constrained variant of the expectation-maximization algorithm~\citep{dempster1977em,baum1970maximization}. Existing approaches to constrained HMM learning have variously employed Lagrangian-dual reformulations~\citep{ma2024structured} and gradient-based penalty methods, but our augmentation-based construction enables the standard EM machinery to be applied directly on the augmented state space, with the constraint enforced by construction at every iteration rather than penalized in the objective. Subsequently, the M-step updates inherit a clean form despite the gating, since the gating annihilates infeasible counts and thus enters the updates only as a support restriction.

\subsubsection{Discrete-Time E-Step and M-Step}
\label{sec:discrete_em}
The E-step requires the augmented smoothed marginals $\gamma_t(i, c)$ defined in Equation~\eqref{eq:gamma}, together with the expected augmented transition counts
\begin{equation}
\xi_t\bigl((i, c),\, (j, c')\bigr) \;=\; \frac{\alpha_t(i, c)\, \tilde{P}_{(i,c),(j,c')}\, B_j(y_{t+1})\, \beta_{t+1}(j, c')}{Z},
\label{eq:xi}
\end{equation}
in which $Z = \Pb(Y_{1:T}, \mathsf{Acc})$ is the constrained marginal likelihood from Equation~\eqref{eq:Z_def} and $\xi_t$ is defined for $t = 0, \ldots, T-1$. Both quantities are computed in a single pass of Algorithm~\ref{alg:constrained_forward_backward}. The M-step subsequently updates the base parameters $(\nu, P, B)$ rather than the augmented kernel $\tilde{P}$ directly, since $\tilde{P}$ is determined by $P$ and the (non-learned) controller specification through Equation~\eqref{eq:live_kernel}. Because the controller update is deterministic and gated transitions contribute zero to the augmented expectations, the base-transition update collapses to
\begin{equation}
\adjustbox{max width=\columnwidth}{$\displaystyle
P_{i,j}^{(k+1)}
\;=\;
\frac{
\sum_{t=0}^{T-1} \sum_{c \in \mathcal{C}}
\xi_t\!\bigl((i, c),\, (j, \tau(c, i, j))\bigr)
\,\mathbf{1}\{F(c, i, j) = 1\}
}{
\sum_{t=0}^{T-1} \sum_{c \in \mathcal{C}} \gamma_t(i, c)
},
$}
\label{eq:Mstep_P}
\end{equation}
in which the indicator $\mathbf{1}\{F(c, i, j) = 1\}$ is operationally redundant (since the corresponding $\xi_t$ is zero by construction whenever the gate fires) but is retained to make the support restriction explicit. The initial-distribution update is standard
\begin{equation}
\nu^{(k+1)}(i) \;=\; \sum_{c \in \mathcal{C}} \gamma_0(i, c),
\label{eq:Mstep_pi}
\end{equation}
and the emission update for the base state retains its unconstrained form except that the smoothed marginals are summed over the controller dimension. The emission updates retain their standard unconstrained form, with the smoothed marginals $\gamma_t(i,c)$ summed over the controller dimension when computing the per-state sufficient statistics. The algorithm iterates equations~\eqref{eq:Mstep_P} and \eqref{eq:Mstep_pi} together with these emission updates and the augmented forward--backward pass until the relative change in the constrained marginal log-likelihood $\log Z$ falls below a prescribed tolerance.

\subsubsection{Continuous-Time M-Step}
\label{sec:cthmm_em}
In the continuous-time setting, the M-step updates the base generator $Q$ rather than a transition matrix, and the requisite expectations involve integrals of matrix exponentials over each inter-observation interval. The closed-form update is reported in Theorem~\ref{thm:ct_em} below, where the derivation proceeds via endpoint-conditioned expectations~\citep{hobolth2005statistical, liu2015efficient}, the auxiliary-matrix identity of~\cite{vanloan1978}, and the augmented forward--backward smoothed pairwise posterior, and is given in full in the Supplementary Materials. 

\begin{theorem}[CT-CHMM constrained EM update]
\label{thm:ct_em}
Given the augmented generator $\bar{Q}$ induced by base generator $Q$ and a deterministic controller specification $(\mathcal{C}, c_0, \tau, F, \mathcal{F}_T)$, let $\bar{n}_{ij}^{(r)}$ and $\bar{\tau}_{i}^{(r)}$ denote the fully-conditioned expected number of base-state transitions $i \to j$ and expected sojourn time in base state $i$ under the smoothed augmented posterior at iterate $Q^{(r)}$. The constrained EM update that maximizes the auxiliary functional $\mathcal{Q}(Q;\, Q^{(r)})$ is
\begin{equation}
Q_{ij}^{(r+1)} \;=\; \frac{\bar{n}_{ij}^{(r)}}{\bar{\tau}_{i}^{(r)}}, \qquad i \neq j,
\label{eq:Mstep_Q}
\end{equation}
with diagonal entries $Q_{ii}^{(r+1)} = -\sum_{j \neq i} Q_{ij}^{(r+1)}$. The resulting iterates ascend the constrained marginal log-likelihood $\log \Pb_{Q}(Y_{0:V},\, \mathsf{Acc})$ monotonically.
\end{theorem}

The update~\eqref{eq:Mstep_Q} is the natural continuous-time analogue of the discrete-time M-step~\eqref{eq:Mstep_P}, with the per-base-pair sufficient statistics $\bar{n}_{ij}^{(r)}$ and $\bar{\tau}_{i}^{(r)}$ replacing the discrete-time augmented transition expectations and the gating again entering only through the augmented kernel via Equation~\eqref{eq:live_kernel}. The dominant computational cost is the matrix exponential of the augmented generator at each distinct inter-observation interval, which is amortized by caching.

\subsubsection{Convergence}
\label{sec:em_convergence}
Standard EM ascends the marginal log-likelihood in expectation under the unconstrained model, but the augmented model differs from this in two structural respects: first, that the augmented kernel is sub-stochastic, with the missing mass corresponding to killed trajectories, and second that the marginal log-likelihood under consideration is the constrained quantity $\log Z = \log \Pb(Y_{1:T}, \mathsf{Acc})$ rather than the unconstrained $\log \Pb(Y_{1:T})$. We therefore record the corresponding convergence guarantee for the constrained EM procedure as a separate result.

\begin{theorem}[Monotone ascent of constrained EM]
\label{thm:em_convergence}
Let $\theta = (\nu, P, B)$ denote the base parameters of a CHMM with fixed controller specification $(\mathcal{C}, c_0, \tau, F, \mathcal{F}_T)$, and let $\theta^{(k)}$ denote the parameter iterate produced by applying the constrained E-step of equations~\eqref{eq:gamma}, \eqref{eq:xi} followed by the M-step of equations~\eqref{eq:Mstep_P}--\eqref{eq:Mstep_Q} starting from $\theta^{(k-1)}$. Then the sequence of constrained marginal log-likelihoods is non-decreasing,
\begin{equation}
\log \Pb_{\theta^{(k+1)}}(Y_{1:T}, \mathsf{Acc}) \;\ge\; \log \Pb_{\theta^{(k)}}(Y_{1:T}, \mathsf{Acc}),
\label{eq:em_ascent}
\end{equation}
with equality if and only if $\theta^{(k+1)}$ is a fixed point of the M-step.
\end{theorem}

We emphasize that the objective in Theorem~\ref{thm:em_convergence} is the joint likelihood of the observations and the feasibility event. Where the sampling mechanism retains only feasible trajectories, the appropriate observed-data likelihood is instead the conditional quantity $\log \Pb_{\theta}(Y_{1:T} \mid \mathsf{Acc}) = \log \Pb_{\theta}(Y_{1:T}, \mathsf{Acc}) - \log \Pb_{\theta}(\mathsf{Acc})$, whose second term depends on $\theta$. The E-step is common to the two objectives, since the latent posterior at fixed $\theta$ is the same, but the M-step is not, and the updates~\eqref{eq:Mstep_P}--\eqref{eq:Mstep_Q} do not in general ascend the conditional likelihood.


\subsection{Computational Complexity}
\label{sec:complexity}
Finally, we record the complexity of constrained inference on the augmented state space.
\begin{theorem}[Complexity of constrained inference]
\label{thm:complexity}
The augmented forward--backward and Viterbi recursions of Algorithms~\ref{alg:constrained_forward_backward} and~\ref{alg:constrained_viterbi} on a CHMM with base state-space cardinality $S$, controller cardinality $|\mathcal{C}|$, and trajectory length $T$ run in time $O(T\,S^{2}\,|\mathcal{C}|)$ in the dense-transition regime, and in time $O(T\,S\,|\mathcal{C}|\,d_{\mathrm{avg}})$ in the sparse regime, where $d_{\mathrm{avg}}$ denotes the average out-degree of the base transition graph. Storage scales as $O(S\,|\mathcal{C}|)$ per time step for forward--backward and $O(T\,S\,|\mathcal{C}|)$ overall for Viterbi.
\end{theorem}

Constrained inference on the augmented state space therefore has the same asymptotic structure as standard hidden Markov model inference on a state space of size $S\,|\mathcal{C}|$, with the controller size $|\mathcal{C}|$ acting as a multiplicative overhead factor relative to the unconstrained cost. In practice, the deterministic controller update and the gating function~$F$ ensure that only a small fraction of the augmented transitions are feasible, with the result that the practical cost is well below the dense-matrix worst case. The precomputed adjacency lists of Section~\ref{sec:sparsity} exploit this sparsity directly.

The controller-overhead factor $|\mathcal{C}|$ is determined entirely by the constraint, and is small for most cataloged families, the all-different family being the exception at $2^{S}$. Moreover, the composition of several controllers multiplies their sizes in the worst case, so that modest overhead is a property of particular specifications and not of the catalog as a whole. Table~\ref{tab:complexity_overhead} of the Supplementary Materials summarizes the overhead factor for the constraint families of Table~\ref{tab:constraint-catalog-complete}.

In the continuous-time setting, the per-interval cost of the recursion in Equation~\eqref{eq:alpha_ct} is dominated by the matrix exponential of the augmented generator, which costs $O\bigl((S\,|\mathcal{C}|)^{3}\bigr)$ per inter-observation interval. The full EM iteration therefore has cost $O(T\,S^{2}\,|\mathcal{C}|)$ in discrete time and $O\bigl(V\,(S\,|\mathcal{C}|)^{3}\bigr)$ in continuous time, where $V$ is the number of intervals.
\section{Robustness to Misspecification}
\label{sec:theoretical}
The framework's exactness, monotone-ascent, and complexity guarantees have all been established as direct consequences of the controller-augmentation construction. In practice, however, the full set of constraints governing any given problem is rarely known in its entirety, and the constraint set elicited from domain knowledge may therefore overlap with but not exactly coincide with the true feasibility set. To this end, in this section we study the behavior of constrained inference under such constraint misspecification.

We establish that constrained inference remains stable under misspecification, with the deviation of the constrained posterior from the truly constrained one bounded by an interpretable quantity depending only on the symmetric difference between the specified and true constraint sets under the unconstrained posterior. The bound concerns posterior distributions, in that the consequences of an incomplete, incorrect or contradictory specification for individual decoded paths are examined empirically in Sections~\ref{sec:completeness-decoding} and~\ref{sec:misspec-decoding}.

\begin{theorem}[Misspecification bound under constraint mismatch]
\label{thm:misspecification}
Let $\chi$ be the specified feasibility set and $\chi^{*}$ the true feasibility set, and fix observations $Y := Y_{1:T}$. Let $\mu(\cdot) := \Pb(X_{0:T} \in \cdot \mid Y)$ denote the unconstrained trajectory posterior, and assume $\mu(\chi) > 0$ and $\mu(\chi^{*}) > 0$. Then the constrained trajectory posteriors satisfy the exact identity
\begin{equation}
\label{eq:tv_exact_events}
\bigl\| \mu(\cdot \mid \chi) - \mu(\cdot \mid \chi^{*}) \bigr\|_{\mathrm{TV}}
\;=\;
1 - \frac{\mu(\chi \cap \chi^{*})}{\max\{\mu(\chi),\, \mu(\chi^{*})\}},
\end{equation}
and the marginal inference error at any time $t$ obeys
\begin{equation}
\label{eq:tv_marginal_bound}
\adjustbox{max width=\columnwidth}{$\displaystyle
\bigl\| \Pb(X_t \mid Y, \chi) - \Pb(X_t \mid Y, \chi^{*}) \bigr\|_{\mathrm{TV}}
\;\le\;
\min\!\left\{1,\; \frac{\mu(\chi \,\triangle\, \chi^{*})}{\max\{\mu(\chi),\, \mu(\chi^{*})\}}\right\},
$}
\end{equation}
where $\triangle$ denotes the set symmetric difference and is defined as $A \, \triangle \, B = (A \cup B) \setminus (A \cap B)$ for sets $A, \, B$. 
\end{theorem}


The bound carries two operational consequences. First, the trajectory-level identity~\eqref{eq:tv_exact_events} states that the constrained posteriors under $\chi$ and $\chi^{*}$ differ in total variation by precisely the fraction of the larger constraint set that is not shared with the smaller one, which is a transparent and interpretable measure of disagreement. Second, the marginal-level bound~\eqref{eq:tv_marginal_bound} states that the per-time-step inference error is at most the relative symmetric difference of the two constraint sets under the unconstrained posterior. The trajectory-level identity attains its maximum of one when the two constraint sets share no posterior mass, but the marginal bound need not be attained in that case, since disjoint sets of trajectories may still induce overlapping one-time marginals, rendering the marginal bound to thus in general be conservative.  Subsequently, the bound depends on the constraint sets only through their unconstrained-posterior masses, and is therefore robust to the specific parametric form of either constraint, requiring only that both retain positive probability under the unconstrained posterior given the observed data.

\section{Evaluation Metrics}
\label{sec:metrics}
Sequential inference is evaluated across application domains using a heterogeneous and often application-specific set of metrics. Position-wise accuracy and macro-averaged $F_{1}$ score remain the standard reporting convention, but neither captures whether a decoded trajectory is structurally valid, temporally well-aligned, or compositionally coherent. In this section, we therefore introduce a small evaluation framework that retains the standard classification metrics whilst supplementing them with measures of trajectory validity and structural fidelity.

Throughout this section we let $\bx \in \mathcal{X}^{T+1}$ denote the ground-truth latent trajectory and $\hat{\bx} \in \mathcal{X}^{T+1}$ a decoded trajectory produced by the inference procedure under evaluation, with constraints summarized by the trajectory feasibility set $\chi \subseteq \mathcal{X}^{T+1}$ introduced in Section~\ref{sec:framework}.

In this section, we report four headline metrics which we will use and report for each experiment we conduct in this paper. 

\textbf{Position-wise accuracy.}
The position-wise accuracy is the pointwise agreement rate
\begin{equation}
\mathrm{Acc}(\bx, \hat{\bx}) \;=\; \frac{1}{T+1} \sum_{t=0}^{T} \mathbf{1}\{\hat{\bx}_{t} = x_{t}\},
\label{eq:acc}
\end{equation}
the standard reporting convention for sequence-labeling tasks~\citep{rabiner1989tutorial}.

\textbf{Macro-averaged $F_{1}$ score.}
For a state label $s \in \mathcal{X}$, write
\[
\mathrm{prec}(s) \;=\; \frac{|\{t : \hat{\bx}_{t} = s,\, x_{t} = s\}|}{|\{t : \hat{\bx}_{t} = s\}|}\]
\[
\mathrm{rec}(s) \;=\; \frac{|\{t : \hat{\bx}_{t} = s,\, x_{t} = s\}|}{|\{t : x_{t} = s\}|}
\]
for the per-label precision and recall, with the convention that either ratio is set to zero when its denominator vanishes. Let $\mathcal{X}_{\mathrm{pres}} = \{s \in \mathcal{X} : s \text{ appears in } \bx \text{ or } \hat{\bx}\}$ denote the set of labels present in the ground-truth or decoded trajectory. The macro-averaged $F_{1}$ score weights every present label equally and is defined as
\begin{equation}
\mathrm{macroF}_{1}(\bx, \hat{\bx}) \;=\; \frac{1}{|\mathcal{X}_{\mathrm{pres}}|} \sum_{s \in \mathcal{X}_{\mathrm{pres}}} \frac{2\, \mathrm{prec}(s)\, \mathrm{rec}(s)}{\mathrm{prec}(s) + \mathrm{rec}(s)},
\label{eq:macrof1}
\end{equation}
with the per-label $F_{1}$ summand set to zero when both precision and recall vanish~\citep{vanrijsbergen1979ir,sokolova2009systematic}.

\textbf{Sequence-level trajectory validity rate.}
The most direct operationalization of the structural argument of this paper is the binary indicator that the entire decoded path lies in the feasibility set
\begin{equation}
\mathrm{TVR}_{\mathrm{seq}}(\hat{\bx}, \chi) \;=\; \mathbf{1}\!\bigl\{\hat{\bx} \in \chi \bigr\},
\label{eq:tvr-seq}
\end{equation}
which evaluates to one when all constraints are met. Aggregated over a test set, $\mathrm{TVR}_{\mathrm{seq}}$ reports the fraction of sequences for which the decoded trajectory is globally valid. By Theorem~\ref{thm:augmented_kernels}, the controller-augmented decoder assigns zero posterior mass to infeasible trajectories, and accordingly achieves $\mathrm{TVR}_{\mathrm{seq}} = 1$ whenever a finite-score feasible path exists and is returned, for any constraint encodable in the catalog of Table~\ref{tab:constraint-catalog-complete}. The empirical role of $\mathrm{TVR}_{\mathrm{seq}}$ is therefore to expose the structural failures of decoders that do not enforce constraints by construction. 

\textbf{Segment-level $F_{1}$ score.}
Position-wise accuracy penalizes label errors but can obscure errors in segment structure, and we therefore complement the previous metrics with a segment-aware $F_{1}$ score that requires both label identity and temporal extent to agree within a prespecified tolerance. To make this precise, define a \emph{segment} of a trajectory $\bx$ to be a triple $(\ell,\, a,\, b) \in \mathcal{X} \times [T] \times [T]$ with $a \le b$ such that $x_{a} = x_{a+1} = \cdots = x_{b} = \ell$ and the run is maximal, in the sense that $x_{a-1} \neq \ell$ when $a > 0$ and $x_{b+1} \neq \ell$ when $b < T$; here $\ell$ is the segment's label, $a$ its start time, and $b$ its end time. Let $\Sigma(\bx)$ denote the set of segments of $\bx$ obtained by run-length encoding, and define $\Sigma(\hat{\bx})$ analogously for the decoded trajectory.

A predicted segment $(\hat{\ell},\, \hat{a},\, \hat{b}) \in \Sigma(\hat{\bx})$ is said to \emph{match} a ground-truth segment $(\ell,\, a,\, b) \in \Sigma(\bx)$ at tolerance $\Delta \ge 0$ if and only if the labels coincide, $\hat{\ell} = \ell$, and the endpoints align within tolerance, $|\hat{a} - a| \le \Delta$ and $|\hat{b} - b| \le \Delta$. Predicted segments are processed in temporal order, whereby each is paired with the eligible unmatched reference segment having the smallest endpoint-distance sum, and where ties follow reference-segment order. Each segment is matched at most once. Subsequently, we let $\mathrm{TP}(\Delta)$ denote the number of matched pairs, $\mathrm{FN}(\Delta) = |\Sigma(\bx)| - \mathrm{TP}(\Delta)$ the number of unmatched ground-truth segments, and $\mathrm{FP}(\Delta) = |\Sigma(\hat{\bx})| - \mathrm{TP}(\Delta)$ the number of unmatched predicted segments. The segment-level $F_{1}$ score is then
\begin{equation}
\mathrm{SegF}_{1}(\bx, \hat{\bx};\, \Delta) \;=\; \frac{2\, \mathrm{TP}(\Delta)}{2\, \mathrm{TP}(\Delta) + \mathrm{FP}(\Delta) + \mathrm{FN}(\Delta)},
\label{eq:segf1}
\end{equation}
which evaluates to 1 when every ground-truth segment is matched by exactly one predicted segment within tolerance, and 0 when no ground-truth segment is matched~\citep{ward2011performance,mesaros2016metrics}.
\section{Experiments}
\label{sec:experiments}
In this section, we evaluate our controller-augmented framework in two parts. We first conduct a series of controlled experiments on synthetic data, which examine the computational and predictive consequences of the framework developed in Sections~\ref{sec:framework} through~\ref{sec:theoretical}. We subsequently evaluate the framework on three real-world sequence labeling tasks of substantively different character: gene-structure decoding in \emph{Drosophila melanogaster}, free-living activity decoding in CASAS smart-home environments, and protocol-defined human activity recognition from wearable inertial sensors. 

For the decoder comparisons, we report means and standard deviations across the natural unit of replication: per-seed for the synthetic experiments, per-home for CASAS, per-gene for Drosophila, and per-subject for HAR. For the real-world neural comparisons, sequences are first averaged within each unit and training seed, then across seeds within unit. Standard deviations describe variation across these units, rather than across all sequence--seed pairs. The Segment-$F_{1}$ tolerance defaults to $\Delta = 0.10\,T$, bur is tightened to $\Delta = 0.05\,T$ on Drosophila where segment boundaries are biologically sharp. The synthetic and real-world comparisons use common training and validation partitions for all methods within each task. 
Implementation details and the per-baseline hyperparameter settings are reported in Supplementary Materials Sections~\ref{app:implementation} and~\ref{app:baselines}. The comparative results comprise five synthetic data-generating seeds and, on each real-world dataset, three neural training seeds on a fixed data split, with the CASAS comparison including all 30 homes for each of the three seeds. Table~\ref{tab:matched-accuracy} in the Supplementary Materials reports paired accuracy differences together with intervals that are conditional on these fits and splits.

\subsection{Numerical Validation on Synthetic Data}
\label{sec:synthetic}
With a fitted HMM held fixed, we first ask how omissions and errors in the constraint specification change the decoded path (Sections~\ref{sec:completeness-decoding} and~\ref{sec:misspec-decoding}), and subsequently compare the controller-augmented decoder against twelve alternative decoders (Section~\ref{sec:comparative-sim}). These experiments use the segment generator described in Supplementary Materials Section~\ref{app:synthetic-dgp}. Section~\ref{sec:recovery-exp} reports a separate learning experiment.

\subsubsection{Constraint Completeness and Decoded Paths}
\label{sec:completeness-decoding}
We first quantify how increasingly complete constraints affect decoded trajectory accuracy and validity. Here, the motivating question is whether partial domain knowledge, corresponding to an incomplete subset of the true constraint rules, provides intermediate benefit, and whether the benefit increases monotonically with constraint completeness.

In this example, for each seed, we hold one fitted HMM fixed and compare six nested specifications: 1) no constraints  2) forbidden edges 3) addition of must-visit rules 4) addition of precedence 5) addition of exactly-one-entry and no-reentry rules 6) and the addition of the stage-order chain. Each setting is evaluated on 30 test sequences, with validity measured against the full generating specification. The last level is logically redundant once the preceding rules are imposed, so the level index measures the listed specification rather than independent amounts of information.
From our experiments, the mean sequence validity rises from $0.873$ without constraints to $0.907$ with must-visit rules, $0.947$ after adding precedence, and $1.000$ after adding the entry-count and no-reentry rules. Subsequently, mean accuracy changes from $0.979$ to $0.981$. In this example, all returned controller paths satisfy the specification actually supplied to the decoder. These results illustrate the benefit of additional correct constraints in this experiment, though they do not establish a general monotonicity result for accuracy or posterior entropy.

\begin{figure}[t]
\centering
\includegraphics[width=0.95\columnwidth]{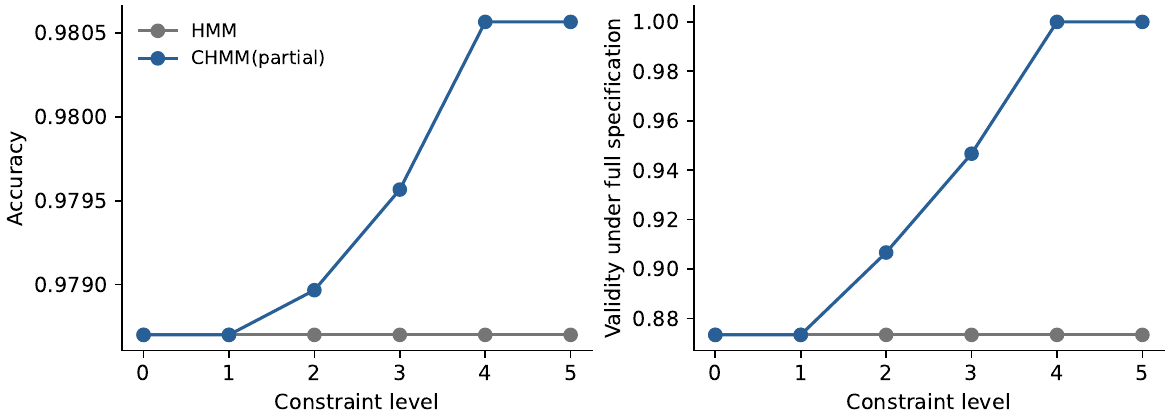}
\caption{Effect of nested constraint specifications on accuracy and validity under the full generating specification, averaged across 5 seeds. The HMM parameters remain fixed within each seed.}
\label{fig:completeness-decoding}
\end{figure}
\subsubsection{Decoding under Omitted, Incorrect and Contradictory Constraints}
\label{sec:misspec-decoding}
This experiment separates omitted constraints from those that may be incorrect or contradictory. The segment generator produces trajectories satisfying the specification $\chi^{*}$. We subsequently decode them under alternative specifications $\chi$.
We hold the fitted HMM fixed and compare four specifications: 1) the correct constraints 2) removal of both precedence and stage-order requirements 3) an incorrect but feasible order obtained by interchanging \textsc{a} and \textsc{b} and 4) contradictory precedence requirements. In the incorrect-order case, incompatible no-skip edges are also removed so that feasible paths remain. In the contradictory case, both activities must occur but each is required to precede the other, so that the feasible set is empty.

Table~\ref{tab:sim-misspecification} distinguishes validity under the supplied specification from validity under the generating specification. Correct constraints provide mean accuracy $0.981$ and true validity $1.000$, while omitting order has little effect and gives true validity $0.993$. On the other hand, the incorrect order still gives used validity $1.000$, but accuracy falls to $0.928$ and true validity to 0. The contradictory specification returns no path on any of the 150 test attempts. These cases distinguish exact satisfaction of the specified rules from accurate recovery of the generating path. 

\begin{table}[t]
\centering
\caption{Constraint misspecification under a fixed fitted HMM, averaged across five seeds (30 sequences per seed). Used validity refers to the supplied constraint; true validity refers to the generating constraint. Contradictory constraints yield no prediction on any of the 150 attempts, so accuracy and conditional used validity are undefined. True validity counts failed attempts as zero.}
\label{tab:sim-misspecification}
\begingroup\fontsize{7.5}{9}\selectfont\setlength{\tabcolsep}{3pt}
\begin{tabular}{lcccc}
\toprule
Specification & Accuracy & Used validity & True validity & Failure rate \\
\midrule
Correct & $0.981$ & $1.000$ & $1.000$ & $0.000$ \\
Order omitted & $0.980$ & $1.000$ & $0.993$ & $0.000$ \\
Incorrect order & $0.928$ & $1.000$ & $0.000$ & $0.000$ \\
Contradictory & --- & --- & $0.000$ & $1.000$ \\
\bottomrule
\end{tabular}
\endgroup
\end{table}

\subsubsection{Comparison Against Baselines}
\label{sec:comparative-sim}
Having examined constraint completeness, we now compare the controller-augmented decoder against alternative inference strategies for constrained sequential models. Specifically, we evaluate 13 decoders on test sequences drawn from a controlled data-generating process whose ground-truth pathwise constraints span the precedence, exactly-$k$, forbidden-edge, and no-reentry families as cataloged in Table~\ref{tab:constraint-catalog-complete}. The synthetic specification uses 5 hidden states arranged around a canonical chain $\textsc{a} \prec \textsc{b} \prec \textsc{c} \prec \textsc{end}$ with a \textsc{rest} state, $K = 8$ observation symbols, and a sequence length of $T = 200$. Here, of 60 development sequences, 48 are used for fitting and 12 for validation, wherein each method is evaluated on 30 held-out test sequences, and with the entire experiment replicated across 5 random seeds.

\textbf{Baselines.}
We deploy 6 baseline decoders alongside our own for comparative evaluation. First, the unconstrained \textsc{hmm} with standard Viterbi decoding~\citep{rabiner1989tutorial} serves as the reference baseline against which the value of any constraint-handling mechanism must be measured. The topology-masked variant, denoted \textsc{tm}, sets the entries of the transition matrix corresponding to forbidden adjacent transitions to zero and re-normalizes row wise, which is the standard pre-controller approach to enforcing local-edge constraints~\citep{durbin1998biological}. The hidden semi-Markov model, denoted \textsc{hsmm}, allows for explicit duration distributions and is the natural baseline for tasks with duration-related structural regularity~\citep{yu2010hidden,hsmm2025recent}. Here, we use the empirical-duration variant fitted from training data, with support extending to the longest observed training run. The structured perceptron, denoted \textsc{sp}, uses linear-chain emission/transition features and averaged perceptron training. The post-hoc filter, denoted \textsc{phf}, computes the top-$50$ Viterbi candidates and returns the highest-scoring constraint-satisfying candidate, providing a finite budget candidate filtering comparison. The beam search decoder with rejection, denoted \textsc{bsr}, performs beam decoding of width $64$ in which transitions to states violating local constraints are pruned at each step, exemplifying the locally pruning strategy that has been deployed in NLP-style structured prediction~\citep{deutsch2019general}. Finally, the controller-augmented \textsc{chmm} (our contribution), augments the hidden state space with a deterministic controller and decodes via Viterbi on an augmented chain. By Theorem~\ref{thm:augmented_kernels}, the \textsc{chmm} decoder assigns zero posterior mass to infeasible trajectories, and returns a feasible optimum whenever a finite-score accepting path exists. 

\textbf{Neural comparators and the isolation of the decoder.} We additionally compare 2 neural encoders with linear-chain CRF heads. The first, denoted \textsc{bilstm}, is a bidirectional LSTM \citep{huang2015bidirectional,lample2016neural}, and the second, denoted \textsc{trf}, is a transformer. These provide learned representations whilst retaining a common form of sequence score.

Each architecture is decoded under 3 rules while its weights are held fixed, so that the decoding rule is the only factor that varies. The unmodified decoder performs unconstrained Viterbi decoding. The $+\textsc{mask}$ variant is a constrained beam decoder of width 32, which checks prefix violations against the full constraint bundle and checks must-visit and exact-count requirements at termination. It can therefore return feasible paths under cumulative constraints, although pruning may remove the optimal path or, indeed, every feasible completion. The $+\textsc{ctrl}$ variant maximizes the same score through the exact controller recursion. All 3 retain the learned unary, pairwise, initial and terminal scores, and controller decoding of a CRF accordingly returns a constrained conditional MAP path.

A direct comparison of CHMMs against the principal alternative strategies (probabilistic model checking, weighted finite-state transducers, posterior regularization, constrained CRFs, HSMMs, and beam-search-with-rejection) is provided in Table~\ref{tab:related-comparison} of the Supplementary Materials, which distinguishes the models, objectives and constraint representations.

Table~\ref{tab:sim-comparative} reports the four headline metrics aggregated across the five seeds. Position-wise accuracy is essentially indistinguishable across the HMM-based decoders, the HSMM and the CHMM agreeing at $0.980$, whilst the CHMM attains the highest macro-$F_1$ and segment-$F_1$, at $0.942$ and $1.000$ respectively. The separation appears instead in sequence-level validity, which is $0.860$ for the unconstrained HMM, $0.940$ for classical beam rejection and $0.987$ for post-hoc filtering, against $1.000$ for the CHMM.

Both neural beam decoders, in common with both exact controller decoders, attain sequence validity $1.000$ on all 150 attempts, so that on this specification the exact recursion confers no feasibility advantage over a beam that checks the full constraint bundle. The difference lies in the path returned. For the BiLSTM, controller decoding raises accuracy from $0.976$ to $0.980$ and segment-$F_1$ from $0.878$ to $0.991$, with the beam attaining $0.979$ and $0.990$ respectively, whereas for the transformer controller, decoding attains accuracy $0.959$ against $0.944$ for the beam and $0.925$ for unconstrained decoding. The predictive benefit of exactness over the beam is thus negligible for the stronger scorer and appreciable for the weaker one, even though the two constrained decoders are identical in observed validity.

\begin{table}[t]
\centering
\caption{Synthetic comparison: mean of five seed-level means, with standard deviation across seeds in parentheses; each seed has 30 test sequences. All methods returned predictions for all 150 attempts. Neural variants share fitted weights within each architecture and seed. \textsc{sp} denotes the structured perceptron.}
\label{tab:sim-comparative}
\begingroup\fontsize{7.5}{9}\selectfont\setlength{\tabcolsep}{3pt}
\begin{tabular}{lcccc}
\toprule
Method & Acc.~$\uparrow$ & macro\,$F_1$~$\uparrow$ & $\mathrm{TVR}_{\mathrm{seq}}$~$\uparrow$ & $\mathrm{SegF}_1$~$\uparrow$ \\
\midrule
\textsc{hmm} & $0.978\,(0.002)$ & $0.930\,(0.012)$ & $0.860\,(0.092)$ & $0.982\,(0.014)$ \\
\textsc{tm} & $0.978\,(0.002)$ & $0.930\,(0.012)$ & $0.860\,(0.092)$ & $0.982\,(0.014)$ \\
\textsc{hsmm} & $0.980\,(0.003)$ & $0.926\,(0.021)$ & $0.867\,(0.088)$ & $0.977\,(0.016)$ \\
\textsc{sp} & $0.971\,(0.002)$ & $0.912\,(0.008)$ & $0.620\,(0.117)$ & $0.944\,(0.014)$ \\
\textsc{phf} & $0.980\,(0.001)$ & $0.940\,(0.004)$ & $0.987\,(0.018)$ & $0.997\,(0.004)$ \\
\textsc{bsr} & $0.979\,(0.001)$ & $0.938\,(0.005)$ & $0.940\,(0.049)$ & $0.992\,(0.009)$ \\
\textsc{bilstm} & $0.976\,(0.003)$ & $0.923\,(0.009)$ & $0.280\,(0.073)$ & $0.878\,(0.023)$ \\
\textsc{bilstm}+\textsc{mask} & $0.979\,(0.002)$ & $0.933\,(0.006)$ & $1.000\,(0.000)$ & $0.990\,(0.004)$ \\
\textsc{bilstm}+\textsc{ctrl} & $0.980\,(0.001)$ & $0.937\,(0.002)$ & $1.000\,(0.000)$ & $0.991\,(0.004)$ \\
\textsc{trf} & $0.925\,(0.006)$ & $0.818\,(0.008)$ & $0.000\,(0.000)$ & $0.445\,(0.051)$ \\
\textsc{trf}+\textsc{mask} & $0.944\,(0.006)$ & $0.844\,(0.011)$ & $1.000\,(0.000)$ & $0.966\,(0.014)$ \\
\textsc{trf}+\textsc{ctrl} & $0.959\,(0.013)$ & $0.889\,(0.033)$ & $1.000\,(0.000)$ & $0.974\,(0.017)$ \\
\textsc{chmm} & $0.980\,(0.001)$ & $0.942\,(0.002)$ & $1.000\,(0.000)$ & $1.000\,(0.001)$ \\
\bottomrule
\end{tabular}
\endgroup
\end{table}

In addition, Figure~\ref{fig:sim-scaling} and Table~\ref{tab:sim-complexity} of the Supplementary Materials report the empirical decoding complexity of the 13 decoders at sequence lengths $T\in\{100,200,400,800\}$.



\subsubsection{Parameter Recovery via Constrained Expectation--Maximization}
\label{sec:recovery-exp}
We now turn to the parameter-estimation regime addressed by Theorem~\ref{thm:em_convergence}, and compare the empirical behavior of constrained and unconstrained EM on data drawn from a feasible population. The reference parameterization $\theta^{*}$ against which both variants are assessed is a supervised fit to feasible trajectories, and not the parameter of the unconstrained chain from which those trajectories were selected, so that the experiment measures distance to that reference and does not address recovery of the pre-selection generating parameters.

In this learning experiment, trajectories satisfying the ground-truth constraint $\chi^{*}$ (precedence, exactly-$k$ visit, forbidden-edge, no-reentry) are drawn by rejection sampling from the unconstrained joint $\Pb_{\theta}(X_{0:T}, Y_{0:T})$, ensuring the generator is independent of the model under evaluation. A reference parameterization $\theta^{*} = (\nu^{*}, P^{*}, B^{*})$ is computed once by supervised maximum likelihood with Laplace smoothing on $n_{\mathrm{ref}} = 500$ constraint-satisfying trajectories, and serves as the recovery target. For each training-set size $N \in \{10, 20, 40, 80, 160\}$ and each of four random seeds, a fresh dataset of $N$ trajectories is generated, the labels are discarded, and both EM variants are initialized identically from $\theta^{(0)} = (1 - \alpha)\,\theta^{*} + \alpha\,\theta_{\mathrm{unif}}$ with $\alpha = 0.30$, holding state identity fixed so that any difference between final estimates is attributable to constraint enforcement rather than to differential alignment. Both variants run standard Baum--Welch and differ only in the state space over which the E-step is computed, with the unconstrained variant on the base HMM, the constrained variant on the augmented chain $\mathcal{X} \times \mathcal{C}$, marginalizing the controller dimension at the M-step to obtain base-level updates for $(\nu, P, B)$. Each runs for at most $30$ iterations with relative-tolerance early stopping at $10^{-5}$.

\begin{figure}[t]
\centering
\includegraphics[width=\columnwidth]{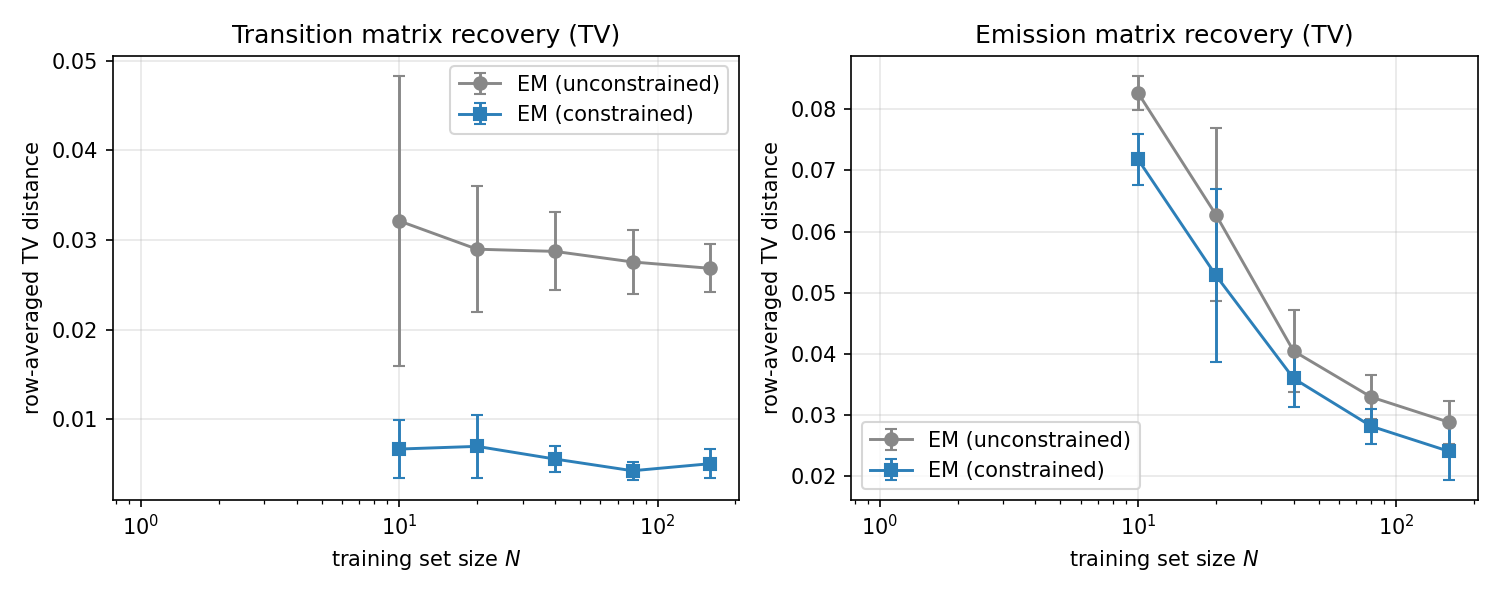}
\caption{Parameter-recovery error of constrained and unconstrained Baum--Welch as a function of training-set size $N$, reported as row-averaged total variation distance to the reference parameters $\theta^{*}$ across 4 random seeds.}
\label{fig:recovery}
\end{figure}

Recovery quality is measured by the row-averaged total variation distance, $\mathrm{TV}(P, P^{*}) = \frac{1}{|\mathcal{X}|}\sum_{i \in \mathcal{X}} \frac{1}{2} \sum_{j \in \mathcal{X}} |P_{ij} - P^{*}_{ij}|$, and analogously for $B$. Results are reported in Figure~\ref{fig:recovery} and Table~\ref{tab:recovery} (Supplementary Materials), and admit a clean separation into two regimes. For the transition matrix, the unconstrained EM exhibits a persistent recovery error of $0.027$--$0.032$ across all training-set sizes, whilst the constrained EM achieves $0.004$--$0.007$ uniformly, a five-fold reduction that does not close as $N$ grows. This persistent difference should be read as a property of the chosen reference and initialization, since the design does not separate the effects of optimization, of selection and of identifiability. We recall, too, that the data here are retained only when feasible, so that the observed-data likelihood for this population is the conditional quantity $\log \Pb_{\theta}(Y \mid \mathsf{Acc})$ of Section~\ref{sec:em_convergence}, whose ascent Theorem~\ref{thm:em_convergence} does not guarantee. For the emission matrix, by contrast, both methods improve with $N$ (from $0.083$ to $0.029$ for the unconstrained variant, $0.072$ to $0.024$ for the constrained), so that the constrained advantage in this regime is a sample-efficiency effect that diminishes as $N$ grows.

The constrained variant also reaches the convergence threshold in roughly 7 iterations across all $N$, whereas the unconstrained variant requires 12--13.

\subsection{Empirical Evaluation on Real-World Data}
\label{sec:realworld}
Having evaluated the framework on controlled synthetic data, we now consider 3 diverse real-world sequence-labeling tasks. The 3 differ in the provenance and character of their constraints: Drosophila carries a biological ordering and entry-count specification, CASAS uses continuous-time constraints mined separately for each home, and HAR uses a training-derived specification for a structured activity recording. Because no-reentry is itself history dependent, these tasks do not partition cleanly into local and cumulative regimes, and so we ask how the supplied constraints, the scoring model and the decoding rule each affect the four headline metrics.

\subsubsection{Constraint Specifications}
\label{sec:mining}
The CHMM framework requires a constraint specification as an input, enabling the evaluated catalogs to be constructed from labeled training data. The Drosophila and CASAS catalogs use the mining procedure below, while HAR uses the separate training derived construction described in Section~\ref{sec:har-exp}. The mining procedure applies a conservative two-stage criterion to the four constraint families (precedence, exactly-$K$-visit cardinality, forbidden-edge, no-reentry), retaining a candidate constraint only when its empirical support on the training trajectories is unity and when the evidence base exceeds a minimum-denominator threshold $\eta_{\mathrm{pres}} = 0.5$ for trajectory-level conditions or $\eta_{\mathrm{edge}} = 10$ for edge-level conditions. The unit-support criterion limits observed training violations but does not guarantee test-time correctness, reflecting the asymmetric cost structure under which a spurious constraint produces systematic test-time error whereas a missed constraint leaves additional trajectories admissible. The full mining procedure with per-family thresholds is given in the Supplementary Materials Section~\ref{app:mining}, and a sensitivity analysis showing that the mined Drosophila catalog is invariant to these thresholds is given in the Supplementary Materials Section~\ref{app:abl-mining}.

\subsubsection{Drosophila Gene-Structure Decoding}
\label{sec:drosophila-exp}


\begin{figure}[t]
\centering
\begin{tikzpicture}[
    cell/.style={draw=black!50, minimum width=0.42cm, minimum height=0.5cm, inner sep=0pt, line width=0.3pt},
    utr5/.style={cell, fill=blue!25},
    start/.style={cell, fill=yellow!55},
    cds/.style={cell, fill=green!30},
    stop/.style={cell, fill=orange!50},
    utr3/.style={cell, fill=red!25},
    rowlabel/.style={anchor=east, font=\small\sffamily},
    legendlabel/.style={anchor=west, font=\small\sffamily},
    scale=0.8
]
\node[rowlabel] at (-0.05, 2.1) {Ground truth};
\foreach \x [count=\i from 0] in
  {utr5,utr5,utr5,utr5,start,cds,cds,cds,cds,cds,cds,cds,stop,utr3,utr3,utr3,utr3} {
  \node[\x] at (0.42*\i + 0.21, 2.1) {};
}
\node[rowlabel] at (-0.05, 1.3) {DT-CHMM (exact)};
\foreach \x [count=\i from 0] in
  {utr5,utr5,utr5,utr5,start,cds,cds,cds,cds,cds,cds,cds,cds,stop,utr3,utr3,utr3} {
  \node[\x] at (0.42*\i + 0.21, 1.3) {};
}
\draw[blue!60!black, dashed, line width=0.7pt]
  (0.42*12, 1.00) rectangle (0.42*14, 1.60);
\node[blue!60!black, font=\scriptsize\sffamily, anchor=south]
  at (0.42*13, 1.62) {boundary shift};
\node[rowlabel] at (-0.05, 0.5) {DT-HMM};
\foreach \x [count=\i from 0] in
  {utr5,utr5,utr5,cds,cds,start,cds,cds,cds,cds,cds,cds,stop,utr3,utr3,utr3,utr3} {
  \node[\x] at (0.42*\i + 0.21, 0.5) {};
}
\draw[red, line width=1pt] (0.42*3, 0.18) rectangle (0.42*6, 0.82);
\node[red, font=\scriptsize\sffamily, anchor=north]
  at (0.42*4.5, 0.16) {precedence violation};
\draw[->, thick] (0, -0.3) -- (0.42*17 + 0.4, -0.3);
\node[font=\small\sffamily] at (0.42*8.5, -0.55) {position $t$};
\foreach \x/\name [count=\i from 0] in {utr5/UTR5, start/START, cds/CDS, stop/STOP, utr3/UTR3} {
  \node[\x] at (2.0*\i - 1.8, -1.2) {};
  \node[legendlabel] at (2.0*\i - 1.6, -1.2) {\name};
}
\end{tikzpicture}
\caption{Schematic illustration of decoder behavior on a \emph{Drosophila} gene transcript. The ground truth (top) follows the canonical biological structure $\textsc{utr5} \prec \textsc{start} \prec \textsc{cds} \prec \textsc{stop} \prec \textsc{utr3}$. DT-CHMM (middle) recovers a feasible segmentation. DT-HMM (bottom) places a CDS segment before START (red box), violating precedence.}
\label{fig:fly-qualitative}
\end{figure}

We employ the annotated \emph{Drosophila melanogaster} genome from FlyBase~\citep{flybase2024}, accessed via the public Kaggle mirror~\citep{oneill2017drosophila}. The task is nucleotide-level sequence labeling with five latent states $\mathcal{X} = \{\textsc{utr5}, \textsc{start}, \textsc{cds}, \textsc{stop}, \textsc{utr3}\}$, restricted to transcripts containing both UTRs and capped at 1000 transcripts before partitioning. The gene-disjoint split yields 560 fitting transcripts from 555 genes, 140 validation transcripts from 138 genes, and 200 evaluated test transcripts from 194 genes. The constraint catalog is mined from the labeled training sequences and recovers the canonical biological structure, with the precedence chain as $\textsc{utr5} \prec \textsc{start} \prec \textsc{cds} \prec \textsc{stop} \prec \textsc{utr3}$, exactly-one-entry cardinalities for all five states, and the 16 biologically invalid forbidden edges (Figure~\ref{fig:fly-qualitative}). Left-to-right edges can prevent backward transitions but do not by themselves require every state to be visited. The full controller also enforces entry counts and terminal acceptance. In this study, the number of controller memory states is distinct from the number of reachable pairs of base and controller states. Table~\ref{tab:drosophila} reports the four headline metrics for the \redit{thirteen} decoders of Section~\ref{sec:comparative-sim} aggregated across the held-out genes.

\begin{table}[t]
\centering
\caption{Drosophila results for 200 transcripts from 194 test genes. Transcripts are averaged within gene and seed, then seeds within gene; entries are means and across-gene standard deviations. Neural seeds are 0, 1 and 2. Predictive scores condition on returned paths. BiLSTM and transformer beams fail on 8 and 28 of 600 attempts, respectively; failures count as zero in validity. Other decoders return all paths.}
\label{tab:drosophila}
\begingroup\fontsize{7.5}{9}\selectfont\setlength{\tabcolsep}{3pt}
\begin{tabular}{lcccc}
\toprule
Method & Acc.~$\uparrow$ & macro\,$F_1$~$\uparrow$ & $\mathrm{TVR}_{\mathrm{seq}}$~$\uparrow$ & $\mathrm{SegF}_1$~$\uparrow$ \\
\midrule
\textsc{hmm} & $0.830\,(0.235)$ & $0.419\,(0.165)$ & $0.642\,(0.479)$ & $0.572\,(0.376)$ \\
\textsc{tm} & $0.831\,(0.235)$ & $0.423\,(0.165)$ & $0.662\,(0.473)$ & $0.572\,(0.377)$ \\
\textsc{hsmm} & $0.743\,(0.319)$ & $0.369\,(0.207)$ & $0.474\,(0.498)$ & $0.500\,(0.404)$ \\
\textsc{sp} & $0.772\,(0.141)$ & $0.224\,(0.060)$ & $0.000\,(0.000)$ & $0.149\,(0.237)$ \\
\textsc{phf} & $0.869\,(0.162)$ & $0.456\,(0.136)$ & $0.781\,(0.413)$ & $0.622\,(0.368)$ \\
\textsc{bsr} & $0.738\,(0.287)$ & $0.315\,(0.154)$ & $0.214\,(0.410)$ & $0.374\,(0.318)$ \\
\textsc{bilstm} & $0.907\,(0.094)$ & $0.493\,(0.080)$ & $0.000\,(0.000)$ & $0.449\,(0.223)$ \\
\textsc{bilstm}+\textsc{mask} & $0.916\,(0.081)$ & $0.533\,(0.096)$ & $0.986\,(0.106)$ & $0.719\,(0.283)$ \\
\textsc{bilstm}+\textsc{ctrl} & $0.917\,(0.079)$ & $0.532\,(0.096)$ & $1.000\,(0.000)$ & $0.718\,(0.286)$ \\
\textsc{trf} & $0.866\,(0.085)$ & $0.449\,(0.062)$ & $0.000\,(0.000)$ & $0.335\,(0.186)$ \\
\textsc{trf}+\textsc{mask} & $0.871\,(0.086)$ & $0.460\,(0.067)$ & $0.953\,(0.134)$ & $0.523\,(0.280)$ \\
\textsc{trf}+\textsc{ctrl} & $0.872\,(0.088)$ & $0.464\,(0.066)$ & $1.000\,(0.000)$ & $0.536\,(0.273)$ \\
\textsc{chmm} & $0.879\,(0.150)$ & $0.488\,(0.112)$ & $1.000\,(0.000)$ & $0.676\,(0.354)$ \\
\bottomrule
\end{tabular}
\endgroup
\end{table}

Among the generative decoders, the CHMM attains mean accuracy $0.879$, macro-$F_1$ $0.488$ and sequence validity $1.000$, against accuracy $0.830$ and validity $0.642$ for the unconstrained HMM. The structured perceptron attains accuracy $0.772$ but macro-$F_1$ of only $0.224$ and segment-$F_1$ of $0.149$, which illustrates that position-wise accuracy alone gives an incomplete account of performance across labels and boundaries. Figure~\ref{fig:fly-qualitative} illustrates the corresponding distinction between a boundary error and a precedence violation.

For the neural scorers, the BiLSTM attains accuracy $0.907$ without constraints, $0.916$ under beam decoding and $0.917$ under controller decoding, with corresponding segment-$F_1$ values of $0.449$, $0.719$ and $0.718$, whilst the transformer attains accuracies of $0.866$, $0.871$ and $0.872$. These predictive summaries are conditional on a path being returned, although the BiLSTM and transformer beams fail to return a feasible terminal path on 8 and 28 of 600 attempts respectively, whereas both exact decoders return feasible paths on all 600. With failures counted as invalid, the gene-weighted beam validity rates are $0.986$ and $0.953$, compared with zero for the unconstrained neural decoders, and all 200 reference paths satisfy the supplied specification.

\subsubsection{CASAS Free-Living Activity Decoding}
\label{sec:casas-exp}
This experiment examines constraints mined from free-living activity recordings, in which adjacent-edge restrictions can be enforced locally whereas no-reentry and stage order require history. CASAS thereby tests whether approximate constrained decoding suffices for the particular specifications that mining produces, without presuming those specifications to be memoryless.

We employ the CASAS smart-home dataset released on Zenodo~\citep{cook2025casas}, which contains longitudinal ambient sensor streams from $189$ community homes collected between $2007$ and $2024$. Within this data, our experiments focus on a subset of $30$ longitudinal apartment recordings from one- and two-resident homes. Across the evaluated homes, the label space contains $12.8$ activity states on average, whilst the observation vocabulary contains $44.8$ symbols per home, including an unknown-symbol category. The raw recordings are continuous-time event streams with irregular inter-event gaps, which we represent as ordered sequences of event times $0 < t_{1} < \cdots < t_{N}$ with corresponding sensor observations.

The constraint catalogue is mined separately for each home using only that home's training data, with each component capped to maintain a tractable controller. Specifically, we retain only the longest stage-order chain (capped at 6 stages), at most 12 forbidden-edge constraints (selected by descending evidence), and at most 8 no-reentry activities (selected by descending presence count). The resulting catalogues have mean stage-chain length $0.87$, mean no-reentry set size $1.57$ and 12 forbidden edges per home. The recorded continuous-time controller product size, $(L+1)2^K$ for chain length $L$ and $K$ no-reentry activities, averages $16.87$ memory states, while activity labels mentioned in constraints are retained even when they exceed the nominal 12-state target.

For each of the $30$ homes, we evaluate 15 decoders on up to five held-out test dates per home, yielding $147$ test sequences in total. The \textsc{ct-hmm}, \textsc{ct-tm}, and \textsc{ct-chmm}\textsc{(greedy)} operate on the native continuous-time event indexed representation, whilst the remaining discrete-time competitors operate on the event indexed sequence with inter-event timing collapsed to unit spacing. Table~\ref{tab:casas} averages days and neural seeds within each home, then assigns equal weight to each home. 

\begin{table}[t]
\centering
\caption{CASAS results for 147 test days from 30 homes. Days are averaged within home and seed, then seeds within home; entries are means and across-home standard deviations. Neural seeds are 0, 1 and 2. All methods return paths. Reference validity is 0.837 with equal home weighting (124 of 147 days satisfy the specification). CT rows use online filtering; event-indexed rows use offline path decoding.}
\label{tab:casas}
\begingroup\fontsize{7.5}{9}\selectfont\setlength{\tabcolsep}{3pt}
\begin{tabular}{lcccc}
\toprule
Method & Acc.~$\uparrow$ & macro\,$F_1$~$\uparrow$ & $\mathrm{TVR}_{\mathrm{seq}}$~$\uparrow$ & $\mathrm{SegF}_1$~$\uparrow$ \\
\midrule
\textsc{ct-hmm} & $0.370\,(0.144)$ & $0.226\,(0.118)$ & $0.173\,(0.327)$ & $0.367\,(0.119)$ \\
\textsc{ct-tm} & $0.370\,(0.144)$ & $0.226\,(0.118)$ & $0.173\,(0.327)$ & $0.367\,(0.119)$ \\
\textsc{ct-chmm}(greedy) & $0.397\,(0.147)$ & $0.230\,(0.120)$ & $1.000\,(0.000)$ & $0.374\,(0.120)$ \\
\textsc{hmm}-\textsc{evt} & $0.310\,(0.122)$ & $0.213\,(0.108)$ & $0.187\,(0.319)$ & $0.415\,(0.111)$ \\
\textsc{tm}-\textsc{evt} & $0.316\,(0.125)$ & $0.218\,(0.110)$ & $0.507\,(0.451)$ & $0.426\,(0.112)$ \\
\textsc{hsmm}-\textsc{evt} & $0.331\,(0.162)$ & $0.241\,(0.164)$ & $0.233\,(0.324)$ & $0.465\,(0.139)$ \\
\textsc{sp}-\textsc{evt} & $0.552\,(0.181)$ & $0.173\,(0.125)$ & $0.813\,(0.344)$ & $0.345\,(0.194)$ \\
\textsc{phf}-\textsc{evt} & $0.311\,(0.123)$ & $0.215\,(0.110)$ & $0.213\,(0.336)$ & $0.416\,(0.112)$ \\
\textsc{bsr}-\textsc{evt} & $0.370\,(0.137)$ & $0.237\,(0.103)$ & $1.000\,(0.000)$ & $0.466\,(0.112)$ \\
\textsc{bilstm}-\textsc{evt} & $0.675\,(0.107)$ & $0.358\,(0.176)$ & $0.842\,(0.262)$ & $0.600\,(0.111)$ \\
\textsc{bilstm}+\textsc{mask}-\textsc{evt} & $0.674\,(0.108)$ & $0.356\,(0.175)$ & $1.000\,(0.000)$ & $0.601\,(0.111)$ \\
\textsc{bilstm}+\textsc{ctrl}-\textsc{evt} & $0.676\,(0.107)$ & $0.360\,(0.176)$ & $1.000\,(0.000)$ & $0.602\,(0.111)$ \\
\textsc{trf}-\textsc{evt} & $0.608\,(0.129)$ & $0.252\,(0.162)$ & $0.648\,(0.353)$ & $0.355\,(0.157)$ \\
\textsc{trf}+\textsc{mask}-\textsc{evt} & $0.607\,(0.129)$ & $0.249\,(0.162)$ & $1.000\,(0.000)$ & $0.354\,(0.157)$ \\
\textsc{trf}+\textsc{ctrl}-\textsc{evt} & $0.609\,(0.128)$ & $0.252\,(0.162)$ & $1.000\,(0.000)$ & $0.358\,(0.156)$ \\
\bottomrule
\end{tabular}
\endgroup
\end{table}

The CASAS labels are heavily imbalanced, with the \textsc{idle} state dominating, and accuracy alone is therefore a poor measure of success. \textsc{sp-evt} has the highest accuracy among the classical rows ($0.552$) but the lowest macro-$F_1$ ($0.173$). The \textsc{ct-chmm}(greedy) attains accuracy $0.397$, macro-$F_1$ $0.230$ and validity $1.000$. The classical beam also attains validity $1.000$, but the two differ in time information, scores and online versus offline inference, so that the fixed-scorer neural comparisons are the more direct test of the decoding rule.

There, the BiLSTM attains accuracy $0.675$, $0.674$ and $0.676$ under unconstrained, beam and controller decoding respectively, and the transformer $0.608$, $0.607$ and $0.609$. Both constrained rules attain validity $1.000$ without failures, against $0.842$ and $0.648$ for the unconstrained models. 

\subsubsection{Human Activity Recognition}
\label{sec:har-exp}
Last, we evaluate on a public wearable-sensor human activity recognition dataset comprising synchronized inertial measurements recorded from body-mounted sensors during a structured multi-activity protocol~\citep{rivas2025multi}. The constraint specification for this task is constructed from the labeled trajectories of the 43 fitting subjects. It records the activity order $1,2,3,4,5$, requires each of these labels to occur without re-entry, requires initial label 0, and forbids state changes absent from the fitting trajectories. The observed compressed label sequence is $0,1,2,0,3,4,0,5,0$, which admits the direct activity transitions $1\to2$ and $3\to4$, so that a resting state is not required between every pair of activities. In this manner, first, the specification is strongly informative, since it very nearly fixes the order of segments and leaves the decoder to locate their boundaries. Second, it is derived from the fitting data, and was fixed only after an initial analysis under an earlier specification proved inconsistent with the released numeric labels.

Sensor observations are multivariate inertial vectors, standardized per channel using a scaler fitted on the training subjects. The decoders compared in this experiment use two emission models. The reference \textsc{hmm-Gaussian}, the topology-masked baseline, and the controller-augmented \textsc{chmm}-Gaussian+\textsc{ctrl} decoder retain the original continuous sensor vectors under a diagonal Gaussian emission model. For the four remaining decoders, denoted \textsc{hsmm-vq}, \textsc{sp-vq}, \textsc{phf-vq}, and \textsc{bsr-vq}, we adopt discrete emission formulations in which the standardized sensor vectors are vector quantized into a $K=64$--symbol alphabet via per-feature equal-frequency binning followed by deterministic hashing of the resulting bin tuple. The \textsc{-vq} suffix marks every decoder operating on the quantized representation. The neural comparators operate on the original continuous vectors and are marked with the \textsc{-cont} suffix. Quantization changes the available representation, so comparisons with the four \textsc{-vq} methods do not isolate the decoder. We therefore interpret the fixed-weight comparisons within each neural architecture separately from comparisons between model families.

\begin{table}[t]
\centering
\caption{HAR results for 13 test subjects. Means and across-subject standard deviations follow averaging over neural seeds 0, 1 and 2. The split contains 43 fitting, 11 validation and 13 test subjects. All reference paths satisfy the training-derived specification. Every decoder returned a path.}
\label{tab:har}
\begingroup\fontsize{7.5}{9}\selectfont\setlength{\tabcolsep}{3pt}
\begin{tabular}{lcccc}
\toprule
Method & Acc.~$\uparrow$ & macro\,$F_1$~$\uparrow$ & $\mathrm{TVR}_{\mathrm{seq}}$~$\uparrow$ & $\mathrm{SegF}_1$~$\uparrow$ \\
\midrule
\textsc{hmm}-Gaussian & $0.840\,(0.138)$ & $0.839\,(0.144)$ & $0.000\,(0.000)$ & $0.197\,(0.089)$ \\
\textsc{tm} & $0.846\,(0.137)$ & $0.847\,(0.144)$ & $0.000\,(0.000)$ & $0.252\,(0.116)$ \\
\textsc{hsmm}-\textsc{vq} & $0.269\,(0.119)$ & $0.190\,(0.126)$ & $0.000\,(0.000)$ & $0.285\,(0.207)$ \\
\textsc{sp}-\textsc{vq} & $0.291\,(0.021)$ & $0.075\,(0.004)$ & $0.000\,(0.000)$ & $0.000\,(0.000)$ \\
\textsc{phf}-\textsc{vq} & $0.282\,(0.039)$ & $0.075\,(0.004)$ & $0.000\,(0.000)$ & $0.014\,(0.050)$ \\
\textsc{bsr}-\textsc{vq} & $0.275\,(0.060)$ & $0.073\,(0.010)$ & $0.000\,(0.000)$ & $0.000\,(0.000)$ \\
\textsc{bilstm}-\textsc{cont} & $0.887\,(0.146)$ & $0.879\,(0.166)$ & $0.179\,(0.322)$ & $0.664\,(0.191)$ \\
\textsc{bilstm}+\textsc{mask}-\textsc{cont} & $0.723\,(0.235)$ & $0.652\,(0.302)$ & $1.000\,(0.000)$ & $0.678\,(0.307)$ \\
\textsc{bilstm}+\textsc{ctrl}-\textsc{cont} & $0.943\,(0.078)$ & $0.943\,(0.084)$ & $1.000\,(0.000)$ & $0.977\,(0.082)$ \\
\textsc{trf}-\textsc{cont} & $0.874\,(0.133)$ & $0.863\,(0.156)$ & $0.051\,(0.125)$ & $0.374\,(0.186)$ \\
\textsc{trf}+\textsc{mask}-\textsc{cont} & $0.656\,(0.230)$ & $0.559\,(0.302)$ & $1.000\,(0.000)$ & $0.598\,(0.299)$ \\
\textsc{trf}+\textsc{ctrl}-\textsc{cont} & $0.923\,(0.093)$ & $0.918\,(0.106)$ & $1.000\,(0.000)$ & $0.957\,(0.107)$ \\
\textsc{chmm}-Gaussian+\textsc{ctrl} & $0.904\,(0.084)$ & $0.906\,(0.087)$ & $1.000\,(0.000)$ & $0.983\,(0.062)$ \\
\bottomrule
\end{tabular}
\endgroup
\end{table}

Table~\ref{tab:har} reports the four headline metrics aggregated across the 13 held-out test subjects. The fixed subject split contains 43 fitting and 11 validation subjects, with neural training seeds 0, 1 and 2, and all three decoding rules for each architecture use the same selected checkpoint within each seed. All 13 reference label sequences satisfy the specification.

Amongst the classical decoders, CHMM-Gaussian+Ctrl is the only method attaining sequence validity $1.000$, with mean accuracy $0.904$ and segment-$F_1$ $0.983$, whereas the Gaussian HMM and its topology-masked variant attain accuracies of $0.840$ and $0.846$ at sequence validity zero. The HSMM attains accuracy $0.269$ and segment-$F_1$ $0.285$ on its quantized representation. Only the Gaussian comparison holds the fitted emission model fixed, the comparisons with the quantized methods changing also the observations available to the decoder.

For the BiLSTM, it gives accuracy $0.887$ under unconstrained decoding, $0.723$ under constrained beam decoding and $0.943$ under exact controller decoding, with both constrained methods attaining validity $1.000$ against $0.179$ for unconstrained decoding. The transformer gives accuracies of $0.874$, $0.656$ and $0.923$ respectively. The beam and the controller therefore satisfy the same specification on every subject and yet return markedly different paths, the beam returning a feasible path that by construction scores no higher than the one returned by the exact recursion, and it is in this sense that exactness, and not merely feasibility, is what the controller contributes on this task. 

In wall-clock terms the BiLSTM beam takes $12.55$ seconds per sequence against $0.246$ seconds for controller decoding, a factor of $51.0$ in the current implementation. These timings include scoring and describe the implementations and not an inherent ordering. The generative CHMM is less accurate than either controller-decoded neural model, although it attains slightly higher segment-$F_1$, so that the ranking depends on both the scoring model and the metric.

\subsection{Cross-Dataset Synthesis}
\label{sec:cross-dataset}
The matched comparisons in Table~\ref{tab:matched-accuracy} hold the neural scores fixed and thereby isolate the decoding rule, and three distinct outcomes emerge. On HAR, exact decoding improves mean subject accuracy over both alternatives for both architectures, even though the beam also returns a valid path for every subject. On gene transcripts, the mean accuracy differences between the two constrained decoders are below $0.1$ percentage point on matched successful attempts, and the decoders are distinguished instead by the beams' failure to return any feasible path on 8 and 28 of 600 attempts for the BiLSTM and transformer respectively. On CASAS, neither accuracy nor reliability separates the two, the main effect of either being validity under a mined specification.

\section{Discussion}
\label{sec:discussion}
We have introduced Controller-Augmented Hidden Markov Models, a framework for exact inference under finite-memory pathwise constraints that transforms a constrained non-Markovian problem into an unconstrained Markovian one on an augmented state space. The augmentation enforces the constraint by construction, which differs qualitatively from constrained-EM methods that project parameters back onto a feasible set in the M-step or impose constraints on the latent posterior in the E-step, and it admits the standard forward--backward, Viterbi, and Baum--Welch machinery on the augmented chain together with a uniformization-based extension to continuous time.

Given a fixed sequence score and at least one feasible path of finite score, controller decoding returns a highest-scoring feasible path, which guarantees feasibility and optimality for the specified objective but not, of course, the highest predictive accuracy.  Agreement with the recorded labels and satisfaction of the supplied constraints are distinct properties, and the two can diverge even when decoding is exact, as the CASAS study shows. Where the specification is correct and strongly informative, as on the wearable-sensor task, exactness is worth much over a beam that satisfies the same constraints. We emphasize that the product of a sequence model with a finite automaton also underlies regular-language constrained CRFs \citep{papay2022regular}, and that the present contribution lies accordingly in the statistical treatment of the generative case and in the controlled comparison, not in the product construction itself.

The framework's principal limitation is the controller state space. Inference cost is polynomial in $|\mathcal{C}|$, but $|\mathcal{C}|$ is determined by the constraint, and not every constraint of practical interest admits a small finite controller. The all-different constraint requires $|\mathcal{C}| = 2^{S}$ at the individual-constraint level, and the combination of multiple non-interacting cumulative constraints multiplies controller sizes in the worst case, although reachability pruning can reduce the compiled state space substantially. A second, separate limitation is the controller-design bottleneck in that, although the catalogue of Table~\ref{tab:constraint-catalog-complete} provides off-the-shelf encodings for eleven constraint families, the design of a correct finite controller for a novel constraint outside the catalogue requires the practitioner to specify $\mathcal{C}$, $\tau$, $F$, and $\mathcal{F}_{T}$ explicitly and to verify the encoding in the sense of Definition~\ref{def:encoding}.

The controller-design bottleneck is in principle addressable through automatic compilation of constraint specifications into minimal finite-state controllers, for which standard automaton translation~\cite{fainekos2006ltl} provides a starting point. We leave the integration of such tooling, and the question of when machine-generated controllers attain small controller sizes, to future work. The controller-size issue motivates structured approximate inference on the augmented state space, for which factored variational approximations exploiting the product structure of $(X_t, C_t)$ in the spirit of factorial HMMs~\citep{ghahramani1997factorial} and beam search on the augmented space (where pruning is more principled than on the base space where the augmented chain is Markovian) are the natural candidates. Beyond these direct extensions, the controller-augmentation principle applies in principle to any state-space model whose unaugmented dynamics are Markovian, with extensions to Constrained Markov Decision Processes~\citep{cmdpsurvey2025} among the possible extensions, while our experiments demonstrate compatibility with neural sequence scores.

\backmatter

\bmhead{Acknowledgements}
This paper describes objective technical results and analysis. Any subjective views or opinions that might be expressed in the paper do not necessarily represent the views of the U.S. Department of Energy or the United States Government.  In preparing this work, the authors used Anthropic's Claude \citep{claude} as
an assistive tool for editorial feedback, consistency checking of the reported
results and references, and generating and formatting several supplementary
tables from author-provided data. The authors reviewed, verified, and
substantially revised all such material and are solely responsible for its
content.

\section*{Declarations}

\bmhead{Funding}
This work was supported by the Laboratory Directed Research and Development program at Sandia National Laboratories, a multi-mission laboratory managed and operated by National Technology and Engineering Solutions of Sandia, LLC, a wholly owned subsidiary of Honeywell International, Inc., for the U.S. Department of Energy's National Nuclear Security Administration under contract DE-NA0003525.



\bmhead{Data availability}
The three real-world datasets are publicly accessible: the
\emph{Drosophila melanogaster} genome annotation at
\url{https://www.kaggle.com/datasets/mylesoneill/drosophila-melanogaster-genome},
the CASAS smart-home activity dataset at
\url{https://zenodo.org/records/15708568}, and the human activity recognition
dataset at \url{https://www.mdpi.com/2306-5729/10/8/129#Data_Description}.

\bmhead{Code availability}
The code implementing all algorithms, baselines, and experimental pipelines
reported in this paper, together with the scripts required to reproduce the
synthetic experiments, will be made publicly available.


\newpage 
\clearpage

\renewcommand{\thesection}{S\arabic{section}}
\renewcommand{\thesubsection}{S\arabic{section}.\arabic{subsection}}
\renewcommand{\thesubsubsection}{S\arabic{section}.\arabic{subsection}.\arabic{subsubsection}}
\renewcommand{\thefigure}{S\arabic{figure}}
\renewcommand{\thetable}{S\arabic{table}}
\renewcommand{\theequation}{S\arabic{equation}}
\renewcommand{\theHsection}{supp.\arabic{section}}
\renewcommand{\theHfigure}{supp.\arabic{figure}}
\renewcommand{\theHtable}{supp.\arabic{table}}
\renewcommand{\theHequation}{supp.\arabic{equation}}
\setcounter{section}{0}
\setcounter{figure}{0}
\setcounter{table}{0}
\setcounter{equation}{0}

\begin{center}
{\Large \bfseries Supplementary Materials\par}
\end{center}
\vspace{1em}

\section{CHMM Examples}
\label{sec:catalog-examples-deferred}
Here, we provide example of three constraints not shown in the main text. 

\subsection{Visitation Constraint}
\label{sec:example_visitation}
In this example, we consider a constraint enforced by the complementary mechanism, namely terminal acceptance. Cardinality constraints of this kind are commonly used in biological sequence analysis to enforce both lower and upper bounds, such as limiting the total number of gap states in an alignment~\citep{christiansen2010inference}, and we illustrate the encoding through a three-state system in which state~$2$ is required to be visited at least once.

\begin{example}[At-Least-One Visitation]
\label{ex:visitation}   
Consider a base HMM with states $\mathcal{X} = \{1, 2, 3\}$ and transition matrix
\begin{equation}
\label{eq:P_visit}
P = \begin{pmatrix}
0.6 & 0.2 & 0.2 \\
0.3 & 0.4 & 0.3 \\
0.2 & 0.3 & 0.5
\end{pmatrix}.
\end{equation}
The constraint $\chi = \{x_{0:T} : \text{state 2 is visited at least once}\}$ is encoded with controller space $\mathcal{C} = \{0, 1\}$ tracking whether state~$2$ has been entered, initialisation $c_0(x) = \mathbf{1}\{x = 2\}$, update $\tau(c, i, j) = c \vee \mathbf{1}\{j = 2\}$, trivial gate $F \equiv 1$, and terminal acceptance $\mathcal{F}_T = \{1\}$.
\end{example}

The constraint here is enforced entirely at termination rather than at any local transition: every base transition is locally admissible, but only those trajectories whose controller flag has been set by time~$T$ are deemed cumulatively feasible. The augmented state graph is shown in Figure~\ref{fig:visitation-graph}, in which transitions that enter state~$2$ for the first time advance the controller from $c = 0$ to $c = 1$ and thereby move the trajectory into the accepting region of the controller, whilst trajectories that fail to enter state~$2$ remain in the rejecting region $c = 0$ and are excluded from the constrained posterior.

\begin{figure}[h]
\centering
\begin{tikzpicture}[
    augstate/.style={rectangle, draw, rounded corners, thick, minimum width=1.5cm, minimum height=1cm},
    acceptstate/.style={rectangle, draw, rounded corners, thick, minimum width=1.5cm, minimum height=1cm, fill=green!20},
    rejectstate/.style={rectangle, draw, rounded corners, thick, minimum width=1.5cm, minimum height=1cm, fill=red!20},
    goodarrow/.style={->, >=stealth, thick},
    scale=0.9
]

\node[rejectstate] (s10) at (0, 3) {$(1,0)$};
\node[rejectstate] (s20) at (4, 3) {$(2,0)$};
\node[rejectstate] (s30) at (8, 3) {$(3,0)$};

\node[acceptstate] (s11) at (0, 0) {$(1,1)$};
\node[acceptstate] (s21) at (4, 0) {$(2,1)$};
\node[acceptstate] (s31) at (8, 0) {$(3,1)$};

\node[gray, above=1.3cm of s20] {$c=0$: not visited 2 (reject at terminal)};
\node[gray, below=1.8cm of s21] {$c=1$: visited 2 (accept at terminal)};

\path[goodarrow] (s10) edge[loop above] node {0.6} (s10);
\path[goodarrow] (s10) edge[bend left=25] node[above] {0.2} (s30);
\path[goodarrow] (s30) edge[bend left=20] node[below] {0.2} (s10);
\path[goodarrow] (s30) edge[loop above] node {0.5} (s30);

\path[goodarrow, blue] (s10) edge[bend right=15] node[left] {0.2} (s21);
\path[goodarrow, blue] (s20) edge[bend left=20] node[right] {0.4} (s21);
\path[goodarrow, blue] (s30) edge[bend left=15] node[right] {0.3} (s21);

\path[goodarrow] (s11) edge[loop below] node {0.6} (s11);
\path[goodarrow] (s11) edge[bend left=20] node[below] {0.2} (s21);
\path[goodarrow] (s11) edge[bend left=25] node[below] {0.2} (s31);

\path[goodarrow] (s21) edge[bend left=20] node[below] {0.3} (s11);
\path[goodarrow] (s21) edge[loop below] node {0.4} (s21);
\path[goodarrow] (s21) edge[bend left=20] node[below] {0.3} (s31);

\path[goodarrow] (s31) edge[bend left=50] node[below] {0.2} (s11);
\path[goodarrow] (s31) edge[bend left=20] node[below] {0.3} (s21);
\path[goodarrow] (s31) edge[loop below] node {0.5} (s31);

\end{tikzpicture}
\caption{Augmented state graph for the at-least-one visitation constraint of state~$2$. Red-shaded states have $c = 0$ and are rejected at terminal time; green-shaded states have $c = 1$ and are accepted. Blue arrows indicate transitions that enter state~$2$ for the first time, switching the controller from $c = 0$ to $c = 1$.}
\label{fig:visitation-graph}
\end{figure}

\subsection{$k$-Segment Constraint}
\label{sec:example_ksegment}

Path constraints restrict the global structure of the trajectory in ways that go beyond ordering or visitation, and we illustrate this through a system with a $k$-segment constraint that fixes the total number of contiguous segments (runs of identical states) along the trajectory. This constraint class was introduced by~\cite{titsias2016statistical}, who proved that an auxiliary counting chain preserves the joint distribution. Here, we show that their construction arises as a special case of our framework, using the controller from the ``$k$-Segment'' row of Table~\ref{tab:constraint-catalog-complete}.

\begin{example}[$k$-Segment Constraint]
\label{ex:ksegment}
Consider a base HMM with states $\mathcal{X} = \{1, 2, 3\}$ and transition matrix
\begin{equation}
\label{eq:P_kseg}
P = \begin{pmatrix}
0.980 & 0.015 & 0.005 \\
0.005 & 0.980 & 0.015 \\
0.015 & 0.005 & 0.980
\end{pmatrix},
\end{equation}
chosen with strong self-transition probabilities to reflect a piecewise-constant signal, as in~\cite{titsias2016statistical}. The constraint $\chi = \{x_{0:T} : \text{the trajectory contains exactly } K \text{ state transitions}\}$ is encoded with controller space $\mathcal{C} = \{0, 1, \ldots, K\}$ counting the number of state transitions observed so far, update $\tau(c, i, j) = c + \mathbf{1}\{i \neq j\}$, gate $F(c, i, j) = 0$ when $i \neq j$ and $c = K$, and terminal acceptance $\mathcal{F}_T = \{K\}$.
\end{example}

A trajectory with $K$ state transitions contains exactly $K + 1$ contiguous segments, and so setting $K = 13$ enforces $14$ segments. The controller increments by one at each state change and blocks further changes once $c = K$, thereby forcing the chain to remain in its current state for the remainder of the sequence, and the terminal condition $\mathcal{F}_T = \{K\}$ further requires that all $K$ transitions were realized, filtering out trajectories with fewer segments. The augmented state space has size $S(K+1) = 3 \times 14 = 42$, and the augmented transition kernel takes the form
\[
\tilde{P}_{(i,c),(j,c')} =
\begin{cases}
P_{ii}\,\mathbf{1}\{c' = c\}, & \text{if } j = i, \\
P_{ij}\,\mathbf{1}\{c < K\}\,\mathbf{1}\{c' = c+1\}, & \text{if } j \neq i,
\end{cases}
\]
in which self-transitions leave the counter unchanged and state changes increment it by one and are blocked once $c = K$. This is precisely the counting-chain augmentation of~\cite{titsias2016statistical}, thus demonstrating that their approach is recovered as a special case of the controller-augmentation framework presented here.

\subsection{Cool-down Constraint}
\label{sec:example_cooldown}
Temporal constraints arise naturally in clinical protocols that mandate minimum spacing between interventions and in manufacturing settings where equipment requires cool-down periods between operations, and we illustrate this through a three-state system with a cool-down period after visiting state~$1$. A related temporal constraint is \emph{minimum duration}, which requires that a new state be occupied for at least $k$ time steps before any transition is permitted, and is used to filter noise in time-series analysis~\citep{Song_2021}.

\begin{example}[Cool-down Period]
\label{ex:cooldown}
Consider a base HMM with states $\mathcal{X} = \{1, 2, 3\}$ and the transition matrix~\eqref{eq:P_visit}. The constraint $\chi = \{x_{0:T} : \text{after leaving state 1, the process cannot re-enter state 1 for } \Delta = 2 \text{ time steps}\}$ is encoded with controller space $\mathcal{C} = \{0, 1, 2\}$ representing a cool-down timer, update function
\[
\tau(c, i, j) =
\begin{cases}
\Delta, & \text{if } i = 1 \text{ and } j \neq 1, \\
\max\{0,\, c - 1\}, & \text{otherwise},
\end{cases}
\]
gate $F(c, i, j) = 0$ when $j = 1$ and $c > 0$, and terminal acceptance $\mathcal{F}_T = \mathcal{C}$.
\end{example}

The cool-down mechanism ensures that, after any visit to state~$1$, the system must wait at least $\Delta = 2$ time steps before re-entering, which suffices to encode minimum spacing requirements that the base chain alone cannot represent. The augmented state space has size $S(\Delta + 1) = 3 \times 3 = 9$, yielding a controller-overhead factor of three.
\section{Proofs of Theoretical Results}
\label{app:proofs}
This appendix collects the proofs of the theoretical results stated in the main text. Specifically, Section~\ref{app:proof-exactness} proves the augmented-kernel result of Theorem~\ref{thm:augmented_kernels}, Section~\ref{app:ct_em_proof} derives the continuous-time EM update of Theorem~\ref{thm:ct_em}, Section~\ref{app:proof-em} proves the constrained Baum--Welch ascent property of Theorem~\ref{thm:em_convergence}, Section~\ref{app:proof-misspec} proves the misspecification bound of Theorem~\ref{thm:misspecification} and Section~\ref{app:proof-complexity} proves the complexity result of Theorem~\ref{thm:complexity}. We adopt the notation of Section~\ref{sec:framework} throughout, with $\tilde{X}_t = (X_t, C_t) \in \tilde{\mathcal{X}}$ denoting the live augmented state, $\bar{X}_t \in \bar{\mathcal{X}} = \tilde{\mathcal{X}} \cup \{\bot\}$ the killed process, $\tau(c, i, j)$ the deterministic controller-update function, $F(c, i, j) \in \{0, 1\}$ the gating function, and $\mathcal{F}_T \subseteq \mathcal{C}$ the terminal-acceptance set.

\subsection{Proof of Theorem~\ref{thm:augmented_kernels}}
\label{app:proof-exactness}
We prove the discrete-- and continuous-time constructions in turn. In each case the argument has two parts: validity of the augmented kernel or generator together with Markovianity of the killed process, followed by the equivalence between the killed augmented chain and the constrained base posterior.

\textbf{Discrete time.}
By construction~\eqref{eq:barP_live}--\eqref{eq:barP_cem} of the main text, $\bar{P}_{s, s'} \ge 0$ for all $s, s' \in \bar{\mathcal{X}}$. For a live state $s = (i, c)$, the row sum then decomposes as
\begin{align}
\sum_{(j, c') \in \tilde{\mathcal{X}}} \bar{P}_{(i,c),\, (j,c')} + \bar{P}_{(i,c),\, \bot}
   &= \sum_{j \in \mathcal{X}} P_{ij}\, F(c, i, j) \notag \\
   &\quad + \biggl(1 - \sum_{j \in \mathcal{X}} P_{ij}\, F(c, i, j)\biggr) \notag \\
   &= 1.
\label{eq:proof-disc-rowsum}
\end{align}
where we have used the determinism of the controller update $c' = \tau(c, i, j)$ to remove the inner sum over $c'$, and the substitution of~\eqref{eq:barP_kill} for the killed mass. The dead row sums to 1 trivially since $\bar{P}_{\bot, \bot} = 1$ and $\bot$ has no other outgoing mass, rendering $\bar{P}$ to therefore be a valid row-stochastic transition matrix on $\bar{\mathcal{X}}$. Since the next state $\bar{X}_{t+1}$ is drawn from $\bar{P}(\cdot \mid \bar{X}_t)$ given only the current state $\bar{X}_t$, the chain $(\bar{X}_t)_{t = 0}^{T}$ is Markovian on $\bar{\mathcal{X}}$ by construction.

For any augmented path $(x_{0:T}, c_{0:T})$ that never visits $\bot$, the path probability under $\bar{P}$ obtained by repeated application of~\eqref{eq:live_kernel} is
\begin{equation}
\resizebox{\columnwidth}{!}{$
p_{0}(x_{0})\, \delta_{c_{0},\, c_{0}(x_{0})}
  \prod_{t = 0}^{T - 1} P_{x_{t},\, x_{t+1}}\, F(c_{t}, x_{t}, x_{t+1})\, \delta_{c_{t+1},\, \tau(c_{t}, x_{t}, x_{t+1})},
  $}
\label{eq:proof-disc-pathprob}
\end{equation}
which equals the unconstrained base path probability $p_{0}(x_{0}) \prod_{t} P_{x_{t}, x_{t+1}}$ multiplied by indicators of local feasibility at every step and of consistency of the controller trace with $\tau$. Because the controller trace is a deterministic function of the base trajectory through $c_{0} = c_{0}(x_{0})$ and $c_{t+1} = \tau(c_{t}, x_{t}, x_{t+1})$, marginalizing~\eqref{eq:proof-disc-pathprob} over $c_{0:T}$ collapses the sum to its single nonzero term, yielding
\begin{equation}
\Pb\!\bigl(X_{0:T} = x_{0:T},\; \mathsf{Acc}\bigr)
\;=\; p_{0}(x_{0}) \prod_{t = 0}^{T - 1} P_{x_{t},\, x_{t+1}}\, \mathbf{1}\{x_{0:T} \in \chi\}.
\label{eq:proof-disc-equiv}
\end{equation}
Here, the indicator follows from Definition~\ref{def:encoding}, specifically the conjunction of pointwise feasibility $F(c_{t}, x_{t}, x_{t+1}) = 1$ for all $0 \le t < T$ together with terminal acceptance $c_{T} \in \mathcal{F}_{T}$, which is by construction equivalent to the global feasibility event $\{x_{0:T} \in \chi\}$. Multiplying by the emission factor $\Pb(Y_{1:T} = y_{1:T} \mid X_{0:T} = x_{0:T})$, which depends on the base trajectory only since the controller does not emit, and conditioning on the feasibility event $\mathsf{Acc}$ yields
\[
p(x_{0:T} \mid y_{1:T},\, \mathsf{Acc}) \;\propto\; p(x_{0:T})\, p(y_{1:T} \mid x_{0:T})\, \mathbf{1}\{x_{0:T} \in \chi\},
\]
which is the constrained posterior of~\eqref{eq:constrained-posterior}.

\textbf{Continuous time.}
For a live state $(i, c)$, all off-diagonal rates in~\eqref{eq:barQ_live}--\eqref{eq:barQ_kill} are nonnegative, and using~\eqref{eq:barQ_diag} the row sum is
\begin{align}
\sum_{\bar{x}' \in \bar{\mathcal{X}}} \bar{Q}_{\bar{x},\, \bar{x}'}
   &= \bar{Q}_{\bar{x},\, \bar{x}} \;+\; \sum_{\bar{x}' \in \tilde{\mathcal{X}},\, \bar{x}' \neq \bar{x}} \bar{Q}_{\bar{x},\, \bar{x}'} \;+\; \bar{Q}_{\bar{x},\, \bot}
   \;=\; 0,
\label{eq:proof-cont-rowsum}
\end{align}
by the definition of $\bar{Q}_{(i,c),\, (i,c)}$ in~\eqref{eq:barQ_diag}, and since the dead row sums to zero trivially. The matrix $\bar{Q}$ is therefore a valid generator. Substituting~\eqref{eq:barQ_live} and~\eqref{eq:barQ_kill} into~\eqref{eq:barQ_diag} further yields
\begin{align*}
\bar{Q}_{\bar{x},\, \bar{x}}
   &= -\sum_{j \neq i} Q_{ij}\, F(c, i, j) \;-\; \sum_{j \neq i} Q_{ij}\, \bigl[1 - F(c, i, j)\bigr] \\
   &= -\sum_{j \neq i} Q_{ij} \;=\; Q_{ii}.
\end{align*}
so that the holding-time distribution in $(i, c)$ is exponential with rate $-Q_{ii}$, which is identical to that of state $i$ in the base chain. Upon a jump from $(i, c)$, the chain transitions to an admissible augmented state $(j, \tau(c, i, j))$ with rate $Q_{ij}$ when $F(c, i, j) = 1$, or is killed (transitions to $\bot$) at total rate $\sum_{j : F(c, i, j) = 0} Q_{ij}$, which is the standard killed-CTMC construction. The holding-time distribution and post-jump distribution at $(i, c)$ depend only on $(i, c)$, so $(\bar{X}_{t})_{t \in [0, T]}$ is a CTMC on $\bar{\mathcal{X}}$ by construction.

For a sample path with jump times $0 = t_{0} < t_{1} < \cdots < t_{N} < T$ and visited base states $x_{0}, \ldots, x_{N}$ together with the deterministically induced controller trace $c_{0}, \ldots, c_{N}$, the path density of the augmented chain restricted to the live state space is
\begin{multline}
p_{0}(x_{0})\, \delta_{c_{0},\, c_{0}(x_{0})}\, \biggl(\prod_{n = 1}^{N} Q_{x_{n-1},\, x_{n}}\, F(c_{n-1}, x_{n-1}, x_{n})\biggr) \\
\times \exp\!\biggl(\sum_{n = 0}^{N} Q_{x_{n}, x_{n}}\, \Delta_{n}\biggr),
\label{eq:proof-cont-pathdensity}
\end{multline}
which equals the unconstrained CTMC path density~\eqref{eq:ctmc_path_density} multiplied by the conjunction of local-feasibility indicators along the trajectory. By Definition~\ref{def:encoding}, restricting to paths that satisfy this conjunction together with terminal acceptance $C_{T} \in \mathcal{F}_{T}$ is equivalent to restricting to $\{x_{0:T} \in \chi\}$, and conditioning on $\mathsf{Acc}$ therefore yields the same characterization of the constrained posterior as in the discrete-time case. \qed

\subsection{Proof of Theorem \ref{thm:ct_em}}
\label{app:ct_em_proof}
Here, we derive the closed-form M-step update of Theorem~\ref{thm:ct_em} in three steps: (i) endpoint-conditioned expectations of the per-interval sufficient statistics; (ii) the smoothed pairwise posterior on augmented endpoints; (iii) the M-step closed form by differentiation of the EM auxiliary functional. The monotone ascent claim then follows from the constrained-EM ascent property of Theorem~\ref{thm:em_convergence} applied to the augmented continuous-time chain.

\textbf{Step (i): endpoint-conditioned expectations.}
For brevity, let $\mathcal{E}_{v} = \{\bar{X}_{t_v} = (i_v, c_v),\, \bar{X}_{t_{v+1}} = (i_{v+1}, c_{v+1})\}$ denote the endpoint event on the $v$-th inter-observation interval, and write $\bar{P}_{v} = \bar{P}(\Delta t_v)_{(i_v, c_v),\, (i_{v+1}, c_{v+1})}$ for the corresponding augmented transition probability. The endpoint-conditioned expected number of base-state transitions $i \to j$ and expected sojourn time in base state $i$ are then, following~\cite{hobolth2005statistical} as adapted to the CT-HMM setting by~\cite{liu2015efficient},
\begin{align}
\Eb\!\bigl[n_{ij} \,\bigl|\, \mathcal{E}_{v};\, Q\bigr]
   &= \frac{Q_{ij}}{\bar{P}_{v}} \int_{0}^{\Delta t_v} \sum_{c}\, \bar{P}(x)_{(i_v, c_v),\, (i, c)} \notag \\
   &\quad \times \bar{P}(\Delta t_v - x)_{(j,\, \tau(c, j)),\, (i_{v+1}, c_{v+1})} \,\mathrm{d}x,
\label{eq:nij_ct_app} \\
\Eb\!\bigl[\tau_i \,\bigl|\, \mathcal{E}_{v};\, Q\bigr]
   &= \frac{1}{\bar{P}_{v}} \int_{0}^{\Delta t_v} \sum_{c}\, \bar{P}(x)_{(i_v, c_v),\, (i, c)} \notag \\
   &\quad \times \bar{P}(\Delta t_v - x)_{(i, c),\, (i_{v+1}, c_{v+1})} \,\mathrm{d}x,
\label{eq:tau_ct_app}
\end{align}
in which the inner sum runs over controller states $c$ reachable from $c_v$ along feasible trajectories, and $\tau(c, j)$ denotes the deterministic controller update following a base-state transition into $j$ from controller state $c$. Both integrals are computed in closed form via the auxiliary-matrix identity of~\cite{vanloan1978}, which establishes that for any matrix $M \in \mathbb{R}^{N \times N}$,
\begin{align*}
\int_{0}^{t} \exp(\bar{Q}\, x)\, M\, \exp\bigl(\bar{Q}\, (t - x)\bigr)\, \mathrm{d}x
   &= \bigl[\exp(A\, t)\bigr]_{1:N,\, N+1:2N}, \\
A &= \begin{pmatrix} \bar{Q} & M \\ 0 & \bar{Q} \end{pmatrix},
\end{align*}
where $N = |\tilde{\mathcal{X}}|$, with $M$ a single-entry matrix isolating the $(i, j)$-entry of $\bar{Q}$ for the transition-count integral and a single-entry matrix isolating the diagonal $(i, i)$-entry for the sojourn-time integral, following~\cite{hobolth2011summary}.

\textbf{Step (ii): smoothed pairwise posterior.}
The fully-conditioned expectations entering the M-step are obtained by averaging the endpoint-conditioned quantities of Step~(i) against the smoothed pairwise posterior on endpoint augmented states. The smoothed pairwise posterior is the natural CT analogue of the discrete-time pairwise posterior $\xi$ of Section~\ref{sec:discrete_em}:
\begin{equation}
\gamma_v(\bar{x}_v;\, \bar{x}_{v+1}) \;=\; \Pb_{Q^{(r)}}\!\bigl(\bar{X}_{t_v} = \bar{x}_v,\, \bar{X}_{t_{v+1}} = \bar{x}_{v+1} \,\bigl|\, Y_{0:V},\, \mathcal{T}_{\mathrm{obs}}\bigr),
\label{eq:gamma_pair_ct_app}
\end{equation}
with the conditioning on $Y_{0:V}$ entering through the augmented forward and backward variables. Specifically, $\gamma_v$ is computed from the augmented forward--backward recursion of Section~\ref{sec:forward_backward} with $\bar{P}$ replaced by the matrix exponential $\bar{P}(\Delta t_v) = \exp(\bar{Q}\,\Delta t_v)$ at each inter-observation interval and the emission factors $B_{i_v}(y_v)$ applied at observation times only, yielding
\begin{equation}
\resizebox{\columnwidth}{!}{$
\gamma_v(\bar{x}_v;\, \bar{x}_{v+1}) \;\propto\; \alpha_v(\bar{x}_v)\, \bar{P}(\Delta t_v)_{\bar{x}_v,\, \bar{x}_{v+1}}\, B_{i_{v+1}}(y_{v+1})\, \beta_{v+1}(\bar{x}_{v+1}),
$}
\label{eq:gamma_pair_factorisation_app}
\end{equation}
with normalization $\Pb(Y_{0:V},\,\mathsf{Acc})$. The fully-conditioned expectation of the per-interval transition count is then
\begin{equation}
\bar{n}_{ij}^{(r)} \;=\; \sum_{v=0}^{V-1} \sum_{\mathcal{E}_v} \gamma_v(\mathcal{E}_v)\, \Eb\!\bigl[n_{ij} \,\bigl|\, \mathcal{E}_v;\, Q^{(r)}\bigr],
\label{eq:nij_full_app}
\end{equation}
and analogously $\bar{\tau}_i^{(r)} = \sum_{v} \sum_{\mathcal{E}_v} \gamma_v(\mathcal{E}_v)\, \Eb[\tau_i \mid \mathcal{E}_v;\, Q^{(r)}]$ for the sojourn time.

\textbf{Step (iii): closed-form M-step.}
The M-step maximizes the EM auxiliary functional
\begin{equation}
\mathcal{Q}(Q;\, Q^{(r)}) \;=\; \Eb_{Q^{(r)}}\!\bigl[\log \Pb_{Q}(\bar{X}_{0:T},\, Y_{0:V},\, \mathsf{Acc}) \,\bigl|\, Y_{0:V},\, \mathcal{T}_{\mathrm{obs}}\bigr],
\label{eq:Q_aux_app}
\end{equation}
which by the standard EM argument lower-bounds the constrained marginal log-likelihood $\log \Pb_{Q}(Y_{0:V},\, \mathsf{Acc})$. The complete-data log-likelihood admits the decomposition
\begin{equation}
\log \Pb_{Q}(\bar{X}_{0:T},\, Y_{0:V},\, \mathsf{Acc}) \;=\; \sum_{i \neq j} \bigl(n_{ij} \log Q_{ij} - \tau_i\, Q_{ij} \bigr) \;+\; \mathrm{const},
\label{eq:complete_loglik_app}
\end{equation}
in which $n_{ij}$ and $\tau_i$ are the trajectory transition counts and sojourn times defined above, and the $Q$-independent terms collect the emission densities, the initial distribution, and the constraint indicator $\mathbf{1}\{\mathsf{Acc}\}$, the last of which depends on $\bar{X}_{0:T}$ but not on $Q$ since the controller dynamics are deterministic. Taking the expectation under the smoothed posterior at $Q^{(r)}$ gives, using the shorthand of Step~(ii),
\begin{equation}
\mathcal{Q}(Q;\, Q^{(r)}) \;=\; \sum_{i \neq j} \bigl( \bar{n}_{ij}^{(r)} \log Q_{ij} \;-\; \bar{\tau}_{i}^{(r)}\, Q_{ij} \bigr) \;+\; \mathrm{const}.
\label{eq:Q_expanded_app}
\end{equation}
Differentiating with respect to $Q_{ij}$ and setting to zero yields the unconstrained-on-the-off-diagonal maximizer
\begin{equation*}
Q_{ij}^{(r+1)} \;=\; \frac{\bar{n}_{ij}^{(r)}}{\bar{\tau}_{i}^{(r)}}, \qquad i \neq j,
\end{equation*}
with the diagonal entries determined by the row-sum constraint $Q_{ii}^{(r+1)} = -\sum_{j \neq i} Q_{ij}^{(r+1)}$.

\textbf{Monotone ascent.}
The constrained CT-CHMM EM procedure is the constrained EM of Section~\ref{sec:em_learning} applied to the killed augmented chain on $\bar{\mathcal{X}}$, with the discrete-time kernel $\tilde{P}$ replaced throughout by the matrix exponential $\bar{P}(\Delta t)$. The monotone-ascent property of Theorem~\ref{thm:em_convergence} therefore applies verbatim, yielding
\[
\log \Pb_{Q^{(r+1)}}(Y_{0:V},\, \mathsf{Acc}) \;\ge\; \log \Pb_{Q^{(r)}}(Y_{0:V},\, \mathsf{Acc}),
\]
with equality if and only if $Q^{(r+1)}$ is a fixed point of the M-step. \qedsymbol

\subsection{Proof of Theorem~\ref{thm:em_convergence}}
\label{app:proof-em}
The proof adapts the standard EM monotone-ascent argument of~\cite{dempster1977em}  to the constrained setting. First, define the EM auxiliary functional
\begin{equation}
\mathcal{Q}(\theta;\, \theta^{(k)}) \;=\; \Eb_{\theta^{(k)}}\!\bigl[\log \Pb(\bar{X}_{0:T},\, Y_{1:T},\, \mathsf{Acc}) \,\bigl|\, Y_{1:T},\, \mathsf{Acc}\bigr],
\label{eq:proof-em-Q}
\end{equation}
in which the expectation is taken with respect to the constrained smoothed posterior $\Pb_{\theta^{(k)}}(\bar{X}_{0:T} \mid Y_{1:T}, \mathsf{Acc})$ at the current iterate $\theta^{(k)}$. The constrained marginal log-likelihood admits the standard decomposition
\begin{equation}
\log \Pb_\theta(Y_{1:T},\, \mathsf{Acc}) \;=\; \mathcal{Q}(\theta;\, \theta^{(k)}) \;-\; \mathcal{H}(\theta;\, \theta^{(k)}),
\label{eq:proof-em-decomp}
\end{equation}
where $\mathcal{H}(\theta;\, \theta^{(k)}) := \Eb_{\theta^{(k)}}[\log \Pb(\bar{X}_{0:T} \mid Y_{1:T},\, \mathsf{Acc}) \mid Y_{1:T},\, \mathsf{Acc}]$, with the decomposition verified by direct substitution of the conditional probability identity inside the expectation.

By Gibbs' inequality applied at the constrained smoothed posterior
\begin{multline}
\mathcal{H}(\theta;\, \theta^{(k)}) - \mathcal{H}(\theta^{(k)};\, \theta^{(k)}) \\
\;=\; -\mathrm{KL}\!\bigl(\Pb_{\theta^{(k)}}(\cdot \mid Y, \mathsf{Acc}) \,\big\|\, \Pb_{\theta}(\cdot \mid Y, \mathsf{Acc})\bigr) \\
\;\le\; 0,
\label{eq:proof-em-jensen}
\end{multline}
with equality if and only if the two constrained smoothed posteriors agree almost surely. Defining the M-step update by $\theta^{(k+1)} \in \arg\max_{\theta} \mathcal{Q}(\theta;\, \theta^{(k)})$ and applying~\eqref{eq:proof-em-decomp} at $\theta^{(k+1)}$ and $\theta^{(k)}$, and $\ell(\theta) := \log \Pb_\theta(Y_{1:T},\, \mathsf{Acc})$ for the constrained marginal log-likelihood under parameters $\theta$, that
\begin{align}
\ell(\theta^{(k+1)}) - \ell(\theta^{(k)})
   &= \bigl[\mathcal{Q}(\theta^{(k+1)};\, \theta^{(k)}) - \mathcal{Q}(\theta^{(k)};\, \theta^{(k)})\bigr] \notag \\
   &\quad - \bigl[\mathcal{H}(\theta^{(k+1)};\, \theta^{(k)}) - \mathcal{H}(\theta^{(k)};\, \theta^{(k)})\bigr] \notag \\
   &\ge 0,
\label{eq:proof-em-final}
\end{align}
since the first bracket is non-negative by the M-step optimization and the second is non-positive by~\eqref{eq:proof-em-jensen}. The monotone-ascent property~\eqref{eq:em_ascent} follows. Equality in~\eqref{eq:proof-em-final} requires both brackets to vanish, which by~\eqref{eq:proof-em-jensen} occurs precisely when the constrained smoothed posterior is unchanged between iterations and $\theta^{(k+1)}$ is a stationary point of $\mathcal{Q}(\cdot;\, \theta^{(k)})$.

The closed-form M-step updates~\eqref{eq:Mstep_pi}, \eqref{eq:Mstep_P}, and~\eqref{eq:Mstep_Q} are obtained by writing the complete-data log-likelihood as a sum of contributions from the initial distribution, transition kernel, and emission distribution, taking the conditional expectation under the constrained smoothed posterior, and maximizing each contribution separately subject to the corresponding simplex constraints. The smoothed augmented marginals $\gamma_{t}(i, c)$ defined in~\eqref{eq:gamma} and pairwise marginals $\xi_{t}(i, c;\, j, c')$ defined in~\eqref{eq:xi} that enter these updates are computed exactly via the augmented forward--backward recursion of Theorem~\ref{thm:augmented_kernels}, which together with the closed-form M-step expressions establishes the tractability of constrained EM. \qed

\subsection{Proof of Theorem~\ref{thm:misspecification}}
\label{app:proof-misspec}
Let $\nu := \mu(\cdot \mid \chi)$ and $\eta := \mu(\cdot \mid \chi^{*})$. Both probability measures are absolutely continuous with respect to the unconstrained posterior $\mu$, with Radon--Nikodym derivatives
\[
\frac{\mathrm{d}\nu}{\mathrm{d}\mu} \;=\; \frac{\mathbf{1}_{\chi}}{\mu(\chi)},
\qquad
\frac{\mathrm{d}\eta}{\mathrm{d}\mu} \;=\; \frac{\mathbf{1}_{\chi^{*}}}{\mu(\chi^{*})}.
\]
Using the standard variational identity $\|\nu - \eta\|_{\mathrm{TV}} = 1 - \int \min\{\mathrm{d}\nu, \mathrm{d}\eta\}$,
\begin{align}
\|\nu - \eta\|_{\mathrm{TV}}
   &= 1 - \int \min\!\biggl\{\frac{\mathbf{1}_{\chi}}{\mu(\chi)},\, \frac{\mathbf{1}_{\chi^{*}}}{\mu(\chi^{*})}\biggr\}\, \mathrm{d}\mu \notag \\
   &= 1 - \frac{\mu(\chi \cap \chi^{*})}{\max\{\mu(\chi),\, \mu(\chi^{*})\}},
\label{eq:proof-misspec-traj}
\end{align}
where the second equality follows from the observation that the integrand is supported on $\chi \cap \chi^{*}$ and equals the smaller of $1/\mu(\chi)$ and $1/\mu(\chi^{*})$ on that set, hence integrates to $\mu(\chi \cap \chi^{*})/\max\{\mu(\chi), \mu(\chi^{*})\}$. This proves the trajectory-level identity~\eqref{eq:tv_exact_events}.

For the marginal-level bound, observe that
\begin{align}
\mu(\chi \,\triangle\, \chi^{*})
   &= \mu(\chi \cup \chi^{*}) - \mu(\chi \cap \chi^{*}) \notag \\
   &\ge \max(\mu(\chi),\, \mu(\chi^{*})) - \mu(\chi \cap \chi^{*}),
\label{eq:proof-misspec-incex}
\end{align}
since $\mu(\chi \cup \chi^{*}) \ge \max\{\mu(\chi),\, \mu(\chi^{*})\}$. Substituting~\eqref{eq:proof-misspec-incex} into~\eqref{eq:proof-misspec-traj} yields
\[
\|\nu - \eta\|_{\mathrm{TV}} \;\le\; \frac{\mu(\chi \,\triangle\, \chi^{*})}{\max\{\mu(\chi),\, \mu(\chi^{*})\}},
\]
which is at most 1 trivially. Finally, the random variable $X_{t}$ is a measurable function of the trajectory $X_{0:T}$, and total variation cannot increase under measurable maps (the data-processing inequality), so
\[
\bigl\| \Pb(X_{t} \mid Y, \chi) - \Pb(X_{t} \mid Y, \chi^{*}) \bigr\|_{\mathrm{TV}}
\;\le\; \|\nu - \eta\|_{\mathrm{TV}},
\]
which delivers the marginal-level bound~\eqref{eq:tv_marginal_bound}. \qed

\subsection{Proof of Theorem~\ref{thm:complexity}}
\label{app:proof-complexity}
The augmented chain has live state space $\tilde{\mathcal{X}} = \mathcal{X} \times \mathcal{C}$ of cardinality $S\, |\mathcal{C}|$. The forward recursion at time $t + 1$ requires computing the augmented forward variable
\[
\bar{\alpha}_{t+1}(j, c') \;=\; B_{j}(y_{t+1}) \sum_{(i, c) \in \tilde{\mathcal{X}}} \bar{\alpha}_{t}(i, c)\, \tilde{P}_{(i, c),\, (j, c')}
\]
for every augmented state $(j, c') \in \tilde{\mathcal{X}}$. A naive implementation would require $O(|\tilde{\mathcal{X}}|^{2}) = O(S^{2}\, |\mathcal{C}|^{2})$ operations per time step, but the augmented kernel~\eqref{eq:live_kernel} is sparse by construction. Specifically, for each augmented source $(i, c)$ and each base destination $j \in \mathcal{X}$, the deterministic controller update produces a unique augmented destination $(j, \tau(c, i, j))$ when $F(c, i, j) = 1$, and produces no augmented destination otherwise. The total number of nonzero entries in the augmented transition kernel is therefore at most $|\tilde{\mathcal{X}}|\, S = S^{2}\, |\mathcal{C}|$, and the recursion requires $O(S^{2}\, |\mathcal{C}|)$ operations per time step. Summing over $T$ time steps yields the dense-regime bound $O(T\, S^{2}\, |\mathcal{C}|)$.

In the sparse base-transition regime where the base transition matrix $P$ has average out-degree $d_{\mathrm{avg}} \ll S$, the number of base destinations $j$ with $P_{ij} > 0$ from each base source $i$ is bounded on average by $d_{\mathrm{avg}}$, so the number of augmented destinations reachable in one step from each augmented source $(i, c)$ is bounded on average by $d_{\mathrm{avg}}$ as well, since each base destination contributes at most one augmented destination through the deterministic controller update. The per-time-step cost reduces to $O(|\tilde{\mathcal{X}}|\, d_{\mathrm{avg}}) = O(S\, |\mathcal{C}|\, d_{\mathrm{avg}})$ and the total cost across $T$ time steps is $O(T\, S\, |\mathcal{C}|\, d_{\mathrm{avg}})$, establishing the sparse-regime bound.

The augmented backward and Viterbi recursions have identical computational structure to the augmented forward recursion, with the inner sum replaced by a maximum in the Viterbi case, and therefore inherit the same complexity bounds. The space complexity is $O(S\, |\mathcal{C}|)$ per time step for the smoothed marginals, and $O(T\, S\, |\mathcal{C}|)$ overall for the Viterbi backpointers. \qed

\section{Algorithms}
In this section, we detail constrained versions of the forward-backward and Viterbi algorithms for utilization in inferring hidden states and parameters in CHMMs. 

\begin{algorithm}[t]
\caption{Constrained forward--backward on the augmented state space.}
\label{alg:constrained_forward_backward}
\begin{algorithmic}[1]
\Require initial distribution $\nu$, base kernel $P$, emission model $B$, controller $(c_0, \tau, F, \mathcal{F}_T)$, observations $y_{1:T}$
\State Initialise $\alpha_0(i, c) \gets \nu(i)\, \mathbf{1}\{c = c_0(i)\}$ for all $(i, c)$
\For{$t = 0, \ldots, T-1$}
    \State $\alpha_{t+1}(j, c') \gets 0$ for all $(j, c')$
    \ForAll{$(i, c) \in \tilde{\mathcal{X}}$ with $\alpha_t(i, c) > 0$, $\;\;j \in \mathcal{X}$ with $F(c, i, j) = 1$}
        \State $c' \gets \tau(c, i, j)$
        \State $\alpha_{t+1}(j, c') \gets \alpha_{t+1}(j, c') + \alpha_t(i, c)\, P_{i,j}\, B_j(y_{t+1})$
    \EndFor
\EndFor
\State Initialise $\beta_T(i, c) \gets \mathbf{1}\{c \in \mathcal{F}_T\}$ for all $(i, c)$
\For{$t = T-1, T-2, \ldots, 0$}
    \State $\beta_t(i, c) \gets 0$ for all $(i, c)$
    \ForAll{$(i, c) \in \tilde{\mathcal{X}}$, $\;\;j \in \mathcal{X}$ with $F(c, i, j) = 1$}
        \State $c' \gets \tau(c, i, j)$
        \State $\beta_t(i, c) \gets \beta_t(i, c) + P_{i,j}\, B_j(y_{t+1})\, \beta_{t+1}(j, c')$
    \EndFor
\EndFor
\State $Z \gets \sum_{i \in \mathcal{X}} \sum_{c \in \mathcal{F}_T} \alpha_T(i, c)$
\If{$Z=0$}\State \Return \redit{infeasible}\EndIf
\State $\gamma_t(i, c) \gets \alpha_t(i, c)\, \beta_t(i, c) / Z$ for all $(t, i, c)$
\State \Return $\bigl\{\sum_{c} \gamma_t(i, c)\bigr\}_{t=0, i \in \mathcal{X}}^{T}$
\end{algorithmic}
\end{algorithm}

\begin{algorithm}[t]
\caption{Constrained Viterbi for MAP inference on the augmented state space.}
\label{alg:constrained_viterbi}
\begin{algorithmic}[1]
\Require initial distribution $\nu$, base kernel $P$, emission model $B$, controller $(c_0, \tau, F, \mathcal{F}_T)$, observations $y_{1:T}$
\State $\delta_0(i, c) \gets -\infty$ for all $(i, c)$
\State $\delta_0\bigl(i, c_0(i)\bigr) \gets \log \nu(i)$ for all $i \in \mathcal{X}$
\For{$t = 0, \ldots, T-1$}
    \State $\delta_{t+1}(j, c') \gets -\infty$ for all $(j, c')$
    \ForAll{$(i, c) \in \tilde{\mathcal{X}}$, $\;\;j \in \mathcal{X}$ with $F(c, i, j) = 1$}
        \State $c' \gets \tau(c, i, j)$;\; $s \gets \delta_t(i, c) + \log P_{i,j} + \log B_j(y_{t+1})$
        \If{$s > \delta_{t+1}(j, c')$}
            \State $\delta_{t+1}(j, c') \gets s$;\; $\psi_{t+1}(j, c') \gets (i, c)$
        \EndIf
    \EndFor
\EndFor
\If{$\max_{(j,c):c\in\mathcal{F}_T}\delta_T(j,c)=-\infty$}\State \Return \redit{infeasible}\EndIf
\State $(x_T^{\star}, c_T^{\star}) \gets \arg\max_{(j, c) : c \in \mathcal{F}_T} \delta_T(j, c)$
\For{$t = T-1, T-2, \ldots, 0$}
    \State $(x_t^{\star}, c_t^{\star}) \gets \psi_{t+1}(x_{t+1}^{\star}, c_{t+1}^{\star})$
\EndFor
\State \Return $x_{0:T}^{\star}$
\end{algorithmic}
\end{algorithm}


\section{Experimental Details}
\label{app:expdetails}

This appendix collects experimental details deferred from Section~\ref{sec:experiments} of the main paper: the synthetic data-generating process, the computational implementation, the baseline tuning protocol, the constraint-mining procedure used for the Drosophila and CASAS experiments, and a qualitative trace illustrating the structural difference between the controller-augmented decoder and the unconstrained baseline on the Drosophila task.

\subsection{Synthetic Data-Generating Process and Parameter Specification}
\label{app:synthetic-dgp}

For the synthetic decoder comparisons of Sections~\ref{sec:completeness-decoding}--\ref{sec:comparative-sim}, the hidden state space is \redit{$\{\textsc{rest},\textsc{a},\textsc{b},\textsc{c},\textsc{end}\}$}, and the observation alphabet has eight symbols. \redit{The generator constructs the sequence $\textsc{rest},\textsc{a},\textsc{rest},\textsc{b},\textsc{rest},\textsc{c},\textsc{rest},\textsc{end}$. Rest durations are integer-truncated Gaussian draws with mean 5 and standard deviation 2, bounded below by 2; activity durations have mean 12 and standard deviation 4, bounded below by 3. The final \textsc{end} segment fills the remaining horizon. Each position emits its state's preferred symbol with probability $0.65$; otherwise a symbol is sampled uniformly from the eight-symbol alphabet. Thus preferred symbols have probability $0.65+0.35/8$.}

\redit{The decoder comparison, nested-rule and incorrect-rule runs use $T=200$, 60 development sequences and 30 test sequences for each of seeds $0,1,2,3,4$. The comparative run splits development sequences into 48 fitting and 12 validation sequences; completeness and misspecification fit one HMM to all 60 development sequences and hold it fixed across specifications. This segment generator is not a rejection-sampled HMM. At $T=200$ its long final segment makes accuracy relatively insensitive to errors in the shorter activities, which motivates reporting macro-$F_1$, segment-$F_1$ and validity alongside accuracy.}

\subsection{Computational Implementation}
\label{app:implementation}

The controller-augmented decoders are implemented as bespoke log-space forward--backward and Viterbi recursions that operate on the augmented transition graph constructed from a mined or protocol-given controller. Standard HMM libraries such as \texttt{hmmlearn} enforce some constraints implicitly through zero entries in the transition matrix, but we observed in preliminary experiments that this approach suffers from numerical-precision issues when the augmented kernel is sparse. Subsequently, our implementation enforces the controller transitions algorithmically through explicit indexing into the augmented adjacency structure, which is both numerically stable and substantially faster on sparse kernels typical of the catalog constraints. Replicates are parallelised across the natural unit of replication using process-level parallelism, with each replicate consuming a single core.

\subsection{Baseline Tuning and Hyperparameter Protocol}
\label{app:baselines}

\redit{The HMM-based categorical decoders share supervised training:} parameters of the underlying HMM (initial distribution, transition matrix, and emission distribution) are estimated by supervised maximum likelihood with Laplace smoothing on the training trajectories, with pseudocount~$0.5$ \redit{for the categorical fits; the HAR Gaussian transition estimate uses $0.01$}, and the controller specification is supplied at decode time only. Baseline-specific hyperparameters are fixed at defensible defaults across all tasks rather than tuned per-task, so that the comparison reflects the structural differences between methods rather than the relative effort spent on hyperparameter search. The HSMM baseline uses an empirical per-state duration distribution \redit{with support extended to the longest observed training run; a fixed short duration cap would force artificial switches in long sequences}. \redit{The structured perceptron is trained over six epochs with emission and transition features and an initial-state bias. Its update is not conditional-likelihood training and it is reported separately from the neural CRFs.} The post-hoc filter searches the top-$K$ Viterbi candidates with $K = 50$, returning the highest-scoring feasible candidate or the top-$1$ path if no feasible candidate is found within the search budget. The beam-search-with-rejection decoder uses beam width~$64$, with rejection applied at every time step to forbidden edges, no-reentry violations, and stage-anchor entries that violate the local stage-order condition. Random seeds for the synthetic experiments are $\{0, 1, 2, 3, 4\}$, and the dataset-specific configuration (training fraction, mining thresholds, vector-quantisation parameters) is reported at the start of each subsection of Section~\ref{sec:realworld} together with the per-task constraint catalog. We do not tune any baseline against test-set performance.

\redit{\textbf{Neural comparators.}
The \textsc{bilstm} comparator is a single-layer bidirectional LSTM with embedding and hidden dimension~$64$ and dropout~$0.1$, followed by a linear-chain CRF head with learnable start, pairwise and end transition scores. The \textsc{trf} comparator is a two-layer transformer encoder with model dimension~$64$, four attention heads, feed-forward dimension~$128$, pre-norm placement, GELU activation, learned positional embeddings and the same linear-chain CRF head, so that the two architectures differ only in their encoder. Symbol-indexed observations enter through an embedding layer with the padding index reserved, whilst continuous observations enter through a linear projection of the standardised feature vector, which permits the wearable-sensor task to be addressed on its native representation rather than through vector quantisation. Both are trained by Adam at learning rate~$10^{-3}$ with weight decay~$10^{-5}$, gradient-norm clipping at~$5.0$ and batch size~$8$, for a maximum of twelve epochs on Drosophila and CASAS and twenty on HAR, with early stopping after three epochs without improvement \redit{in held-out validation token negative log-likelihood, restoring the best checkpoint}. The $+\textsc{mask}$ decode uses beam width~$32$.}

\redit{For each architecture and training seed, a single network is fitted and decoded under all three rules. These comparisons control the scoring model and its fitted weights. They do not remove the need for repeated training seeds when estimating how the effect generalizes across fitted models.}

\redit{The neural comparators are not exhaustively tuned, so their reported performance does not establish a comparison against the strongest attainable sequence model. Long sequences are scored in non-overlapping chunks (length 512 on HAR, 2,048 on Drosophila and 1,024 on CASAS), limiting encoder context; the concatenated scores are decoded as one sequence. All three real-world datasets use neural training seeds 0, 1 and 2 on a fixed split. The reported across-unit variation and paired intervals condition on those fits; they do not quantify uncertainty from new training samples or alternative splits. These limitations can affect predictive accuracy and the paths retained by beam search. Exact controller feasibility, conditional on a finite-score feasible path, is a structural guarantee rather than a consequence of tuning.}

\redit{For the paired comparisons in Table~\ref{tab:matched-accuracy}, predictions are matched by sequence and training seed before differences are averaged within subject, gene or home. Intervals resample these units, preserving all matched predictions within each unit. When a beam returns no path, its predictive metrics are undefined and the paired predictive comparison omits that attempt; validity and the reported failure count retain it.}

\begin{table}[t]
\centering
\caption{Paired accuracy differences in percentage points for exact controller decoding. Intervals are pointwise 95\% percentile bootstrap intervals from 4,000 resamples of subjects, genes or homes, conditional on the fitted models and fixed split. Matching precedes averaging within unit and training seed. Predictive comparisons exclude failed beam attempts; the excluded-attempt column gives their number. These intervals do not quantify training-data or split uncertainty and are not simultaneous intervals.}
\label{tab:matched-accuracy}
\begingroup\fontsize{7.5}{9}\selectfont\setlength{\tabcolsep}{3pt}
\begin{tabular}{lcccccc}
\toprule
Dataset & Scorer & Difference & Mean & 95\% interval & Units & Excluded \\
\midrule
HAR & BiLSTM & Ctrl minus plain & $5.60$ & $[1.84,\,10.72]$ & 13 & 0 \\
HAR & BiLSTM & Ctrl minus beam & $22.08$ & $[12.19,\,31.76]$ & 13 & 0 \\
HAR & Transformer & Ctrl minus plain & $4.90$ & $[2.31,\,8.22]$ & 13 & 0 \\
HAR & Transformer & Ctrl minus beam & $26.62$ & $[17.08,\,36.65]$ & 13 & 0 \\
Drosophila & BiLSTM & Ctrl minus plain & $0.99$ & $[0.46,\,1.73]$ & 194 & 0 \\
Drosophila & BiLSTM & Ctrl minus beam & $0.09$ & $[-0.09,\,0.29]$ & 192 & 8 \\
Drosophila & Transformer & Ctrl minus plain & $0.55$ & $[0.35,\,0.77]$ & 194 & 0 \\
Drosophila & Transformer & Ctrl minus beam & $0.04$ & $[-0.16,\,0.25]$ & 194 & 28 \\
CASAS & BiLSTM & Ctrl minus plain & $0.054$ & $[0.005,\,0.141]$ & 30 & 0 \\
CASAS & BiLSTM & Ctrl minus beam & $0.157$ & $[-0.113,\,0.553]$ & 30 & 0 \\
CASAS & Transformer & Ctrl minus plain & $0.050$ & $[-0.004,\,0.121]$ & 30 & 0 \\
CASAS & Transformer & Ctrl minus beam & $0.182$ & $[0.001,\,0.481]$ & 30 & 0 \\
\bottomrule
\end{tabular}
\endgroup
\end{table}

\subsection{Constraint-Mining Procedure for the Drosophila and CASAS Experiments}
\label{app:mining}

The mining procedure operates on segment-compressed labeled training trajectories and applies a conservative two-stage criterion, in which a candidate constraint is retained only when its empirical support on the training trajectories is unity (the constraint is satisfied in every training instance for which it is defined) and when the supporting evidence base meets a minimum-denominator threshold. The unit-support criterion rejects observed training violations, but finite-sample support does not establish a true population constraint, reflecting the asymmetric cost structure of constrained inference, in that a spurious constraint produces systematic structural error on every test trajectory whose ground truth violates it, whereas a missed constraint leaves a larger admissible search space.

We mine four constraint families.

\textbf{Precedence mining.}
For each ordered pair of distinct labels $(a, b)$ in the state space, we record the fraction of training trajectories in which $b$ appears and in which the first occurrence of $a$ precedes the first occurrence of $b$, and we retain the precedence relation $a \prec b$ when this fraction equals 1. To guard against spurious high-support estimates derived from few observations, we additionally require that $b$ appears in a fraction $\eta_{\mathrm{pres}} = 0.5$ of the training trajectories. When the resulting precedence relation is transitively redundant (in the sense that there exists an alternative path $a \to \cdots \to b$ via other retained relations), the redundant edge is removed.

\textbf{Visit-count mining.}
For each label $a \in \mathcal{X}$, we collect the entry counts of $a$ across the training trajectories, where an entry corresponds to the start of a maximal contiguous segment with label $a$. When the entry count is constant across every training trajectory and equals some value $K \le K_{\max} = 5$, we retain the exactly-$K$-visit cardinality constraint for label $a$.

\textbf{Forbidden-edge mining.}
For each ordered pair of distinct labels $(a, b)$, we record the count of adjacent transitions $a \to b$ across the segment-compressed training trajectories, together with the total number of transitions out of label $a$. When the transition count is zero and the denominator exceeds an evidence threshold $\eta_{\mathrm{edge}} = 10$, we retain the forbidden-edge constraint $a \nrightarrow b$.

\textbf{No-reentry mining.}
For each label $a \in \mathcal{X}$, we record the fraction of training trajectories in which $a$ appears and in which it appears as a single contiguous segment (so that no reentry into $a$ occurs). We retain the no-reentry constraint for $a$ when this fraction equals 1 and when $a$ appears in at least $\eta_{\mathrm{edge}}$ training trajectories.

Following the mining procedure, the per-family constraint sets are composed into a single controller via the catalogue encodings of Section~\ref{sec:catalog}. Sensitivity of the resulting catalogue size and headline metrics to the threshold choices is reported in Appendix~\ref{app:abl-mining}, for the Drosophila task.

\section{Ablation and Sensitivity Analyses}
\label{app:ablation}

This section reports a sensitivity analysis for the constraint-mining thresholds used in the Drosophila experiment of Section~\ref{sec:drosophila-exp}, the objective being to verify that the mined catalog, and with it the decoded paths, does not depend on the particular threshold values chosen.

\subsection{Sensitivity to Mining Thresholds (Drosophila)}
\label{app:abl-mining}

The constraint-mining procedure is governed by 3 thresholds: the precedence-evidence threshold $\eta_{\mathrm{pres}}$, the visit-count cardinality cap $K_{\max}$, and the forbidden-edge evidence threshold $\eta_{\mathrm{edge}}$. We use defaults $\eta_{\mathrm{pres}} = 0.5$, $K_{\max} = 5$, and $\eta_{\mathrm{edge}} = 10$ throughout this sensitivity analysis. To assess sensitivity to these choices, we vary 1 threshold at a time, holding the other two at their default values, and report the size of the resulting mined catalog together with the 4 headline metrics achieved by the controller-augmented decoder. The thresholds act on the label trajectories of the fitting transcripts alone. Wherever the mined catalog coincides with that of the headline comparison, therefore, the controller and the decoded paths coincide with it also, and the metrics are those of the \textsc{chmm} row of Table~\ref{tab:drosophila}.

The CASAS dataset is excluded from this ablation because the per-home heterogeneity makes it difficult to disentangle the threshold-sensitivity effect from the home-specific catalog variation. Instead, the Drosophila dataset is the natural mining-sensitivity example since the underlying biological grammar provides a fixed ground-truth catalog against which the mining outputs may be compared.

\textbf{Precedence-evidence threshold.}
We sweep $\eta_{\mathrm{pres}} \in \{0.1, 0.3, 0.5, 0.7, 0.9\}$ and report the number of retained precedence relations after transitive reduction together with the headline metrics in Table~\ref{tab:abl-eta-pres}. The mining procedure recovers the canonical 4-relation chain $\textsc{utr5} \prec \textsc{start} \prec \textsc{cds} \prec \textsc{stop} \prec \textsc{utr3}$ at every threshold value in the sweep range. This unanimous recovery reflects the structural property that every transcript retained after both-UTRs filtering contains all 5 latent states, whereby the empirical fraction on which the threshold acts equals 1 for every candidate relation and the threshold is non-binding within $[0.1, 0.9]$. Subsequently, the headline metrics are constant across the sweep, which constitutes a strong robustness finding: the recovery of the canonical biological grammar does not depend on a particular choice of precedence-evidence threshold within an order of magnitude of the default.

\begin{table}[h]
\centering
\caption{Sensitivity of the Drosophila mined catalog and decoding metrics to the precedence-evidence threshold $\eta_{\mathrm{pres}}$. The default is $\eta_{\mathrm{pres}} = 0.5$. Reported metrics are the gene-level means of Table~\ref{tab:drosophila}.}
\label{tab:abl-eta-pres}
\begingroup\footnotesize
\begin{tabular}{cccccc}
\toprule
$\eta_{\mathrm{pres}}$ & $|\mathcal{P}|$ (post-reduction) & Acc. & macro\,$F_{1}$ & $\mathrm{TVR}_{\mathrm{seq}}$ & $\mathrm{SegF}_{1}$ \\
\midrule
$0.1$ & $4$ & $0.879$ & $0.488$ & $1.000$ & $0.676$ \\
$0.3$ & $4$ & $0.879$ & $0.488$ & $1.000$ & $0.676$ \\
$\mathbf{0.5}$ & $4$ & $0.879$ & $0.488$ & $1.000$ & $0.676$ \\
$0.7$ & $4$ & $0.879$ & $0.488$ & $1.000$ & $0.676$ \\
$0.9$ & $4$ & $0.879$ & $0.488$ & $1.000$ & $0.676$ \\
\bottomrule
\end{tabular}
\endgroup
\end{table}

\textbf{Visit-count cardinality cap.}
We sweep $K_{\max} \in \{2, 3, 5, 10, 20\}$ and report the number of retained cardinality constraints together with the headline metrics in Table~\ref{tab:abl-Kmax}. The mining procedure recovers the 5 exactly-1-entry cardinality constraints corresponding to the 5 latent states at every value in the sweep range. This invariance reflects the structural property that every state appears exactly once in every retained training transcript, whereby the cap $K_{\max}$ is non-binding for any value $\geq 1$. The headline metrics are therefore constant across the sweep, and the recovery of the cardinality structure is robust across an order of magnitude of variation in the cap.

\begin{table}[h]
\centering
\caption{Sensitivity of the Drosophila mined catalog and decoding metrics to the visit-count cap $K_{\max}$. The default is $K_{\max} = 5$.}
\label{tab:abl-Kmax}
\small
\begin{tabular}{cccccc}
\toprule
$K_{\max}$ & $|\mathcal{V}|$ & Acc. & macro\,$F_{1}$ & $\mathrm{TVR}_{\mathrm{seq}}$ & $\mathrm{SegF}_{1}$ \\
\midrule
$2$ & $5$ & $0.879$ & $0.488$ & $1.000$ & $0.676$ \\
$3$ & $5$ & $0.879$ & $0.488$ & $1.000$ & $0.676$ \\
$\mathbf{5}$ & $5$ & $0.879$ & $0.488$ & $1.000$ & $0.676$ \\
$10$ & $5$ & $0.879$ & $0.488$ & $1.000$ & $0.676$ \\
$20$ & $5$ & $0.879$ & $0.488$ & $1.000$ & $0.676$ \\
\bottomrule
\end{tabular}
\end{table}

\textbf{Forbidden-edge evidence threshold.}
We sweep $\eta_{\mathrm{edge}} \in \{1, 5, 10, 20, 50\}$ and report the number of retained forbidden edges together with the headline metrics in Table~\ref{tab:abl-eta-edge}. The mining procedure recovers the 16 biologically invalid pairwise transitions at every threshold value in the sweep range. The 5-state space exhibits $5 \times 4 = 20$ ordered pairs of distinct states, of which 4 form the canonical chain and the remaining 16 are forbidden under the canonical grammar. Since each transition out of any state is observed in well over 50 training transcripts (the smallest training-set evidence base exceeds the largest threshold tested), the threshold is non-binding within $[1, 50]$. The headline metrics are therefore constant across the sweep.

\begin{table}[h]
\centering
\caption{Sensitivity of the Drosophila mined catalog and decoding metrics to the forbidden-edge evidence threshold $\eta_{\mathrm{edge}}$. The default is $\eta_{\mathrm{edge}} = 10$.}
\label{tab:abl-eta-edge}
\small
\begin{tabular}{cccccc}
\toprule
$\eta_{\mathrm{edge}}$ & $|\mathcal{F}|$ & Acc. & macro\,$F_{1}$ & $\mathrm{TVR}_{\mathrm{seq}}$ & $\mathrm{SegF}_{1}$ \\
\midrule
$1$ & $16$ & $0.879$ & $0.488$ & $1.000$ & $0.676$ \\
$5$ & $16$ & $0.879$ & $0.488$ & $1.000$ & $0.676$ \\
$\mathbf{10}$ & $16$ & $0.879$ & $0.488$ & $1.000$ & $0.676$ \\
$20$ & $16$ & $0.879$ & $0.488$ & $1.000$ & $0.676$ \\
$50$ & $16$ & $0.879$ & $0.488$ & $1.000$ & $0.676$ \\
\bottomrule
\end{tabular}
\end{table}

\textbf{Discussion.}
The 3 Drosophila sensitivity tables collectively establish that the recovery of the canonical biological structure is structurally invariant to the choice of the mining threshold within the ranges considered. Whilst this invariance may at first reading appear to constitute an absence of sensitivity, it is in fact the substantive finding the analysis was designed to elicit, namely that the mining procedure is robust to threshold choice on a dataset whose underlying constraint structure is consistent across training trajectories. The $\mathrm{TVR}_{\mathrm{seq}} = 1$ guarantee is preserved unconditionally across the sweep, the catalog sizes ($|\mathcal{P}| = 4$, $|\mathcal{V}| = 5$, $|\mathcal{F}| = 16$) are unchanged across an order of magnitude of variation in each threshold, and the headline metrics are correspondingly stable. 

\section{Tables}

This section provides formal tables addressed in the main text. 

\begin{table*}[h]
\centering
\caption{Mathematical notation and symbols used throughout this paper.}
\label{tab:notation}
\resizebox{\textwidth}{!}{
\begin{tabular}{ll}
\toprule
\textbf{Symbol} & \textbf{Description} \\
\midrule
$\mathcal{X} = \{1, \ldots, S\}$ & Finite base state space of size $S$ \\
$\mathcal{C}$ & Controller state space (constraint-dependent) \\
$\tilde{\mathcal{X}} = \mathcal{X} \times \mathcal{C}$ & Augmented (live) state space \\
$\bot$ & Dead (killed) state \\
$\bar{\mathcal{X}} = \tilde{\mathcal{X}} \cup \{\bot\}$ & Extended state space including dead \\
$\mathcal{T}$ & Time index set: $\{0,\ldots,T\}$ (discrete) or $[0,T]$ (continuous) \\
$\chi \subseteq \mathcal{X}^{\mathcal{T}}$ & Constraint set (feasible trajectories) \\
$X_t, C_t, \tilde{X}_t, \bar{X}_t$ & Base, controller, augmented, and killed states at time $t$ \\
$P = (P_{ij})$ & Base transition matrix (discrete time) \\
$Q = (Q_{ij})$ & Base generator matrix (continuous time) \\
$\tilde{P}$ & Sub-stochastic live augmented kernel on $\tilde{\mathcal{X}}$ \\
$\bar{P}$ & Stochastic killed kernel on $\bar{\mathcal{X}}$ \\
$\bar{Q}$ & Killed generator on $\bar{\mathcal{X}}$ \\
$B_x(\cdot)$ & Emission distribution for base state $x$ \\
$c_0\colon \mathcal{X} \to \mathcal{C}$ & Controller initialization function \\
$\tau : \mathcal{C}\times \mathcal{X}\times \mathcal{X}\times \mathcal{T} \to \mathcal{C}$ & Controller update function \\
$F : \mathcal{C}\times \mathcal{X}\times \mathcal{X}\times \mathcal{T} \to \{0,1\}$ & Feasibility (gating) function \\
$\mathcal{F}_T \subseteq \mathcal{C}$ & Terminal acceptance set \\
$\mathbf{1}_A(x)$ & Indicator: $1$ if $x \in A$, and $0$ otherwise \\
$\varepsilon_A(i,j)$ & Entry indicator: $\mathbf{1}\{i \notin A,\, j \in A\}$ \\
$k\colon \mathcal{X} \to \{0,\ldots,G-1\}$ & Stage mapping function \\
$\mathcal{B} \subseteq \mathcal{X} \times \mathcal{X}$ & Set of forbidden transitions \\
$a \lor b,\; a \land b$ & Logical OR and AND \\
$\mathsf{Acc}$ & Feasibility event: $\{\bar{X}_t \neq \bot\;\forall\, t \le T\} \cap \{C_T \in \mathcal{F}_T\}$ \\
\bottomrule
\end{tabular}
}
\end{table*}

\begin{table}[h]
\centering
\caption{Controller-overhead factor $|\mathcal{C}|$ relative to the unconstrained HMM cost, for the constraint families of Table~\ref{tab:constraint-catalog-complete}. The constrained inference cost is the unconstrained cost multiplied by $|\mathcal{C}|$ in the worst case, and is typically smaller in practice owing to the sparsity of the gated transition matrix.}
\label{tab:complexity_overhead}
\begin{tabular}{ll}
\toprule
\textbf{Constraint family} & \textbf{Overhead $|\mathcal{C}|$} \\
\midrule
Forbidden edges, no-dwell & $1$ \\
Precedence $a \prec b$ & $2$ \\
Stage monotone with $G$ stages & $G$ \\
At-least-$K$, at-most-$K$, exactly-$K$ visits & $K + 1$ \\
$K$-transition (so $K + 1$ segments) & $K + 1$ \\
No-reentry & $3$ \\
Cool-down with timer $\Delta$ & $\Delta + 1$ \\
All-different on $S$ states & $2^{S}$ \\
\bottomrule
\end{tabular}
\end{table}


\begin{table*}[t!]
\centering
\caption{Methodological comparison of Controller-Augmented Hidden Markov Models against principal alternative strategies for constrained sequential inference. ``Local'' refers to memoryless constraints such as forbidden edges or no-dwell rules, whereas ``cumulative'' refers to constraints carrying path memory, including precedence chains, exactly-$K$-visit cardinalities, and no-reentry. ``Exact'' indicates that the constrained posterior or maximum a posteriori path is computed without sampling, beam, or variational approximation. The learning column identifies the associated estimation procedure, with CT denoting continuous time. The complexity column reports the controller-size overhead relative to the unconstrained inference cost at fixed catalogue size.}
\label{tab:related-comparison}
\begingroup\fontsize{8}{10}\selectfont\setlength{\tabcolsep}{3pt}
\begin{tabularx}{\textwidth}{@{}>{\raggedright\arraybackslash}p{.22\textwidth}>{\raggedright\arraybackslash}p{.13\textwidth}>{\raggedright\arraybackslash}p{.12\textwidth}>{\raggedright\arraybackslash}p{.13\textwidth}>{\raggedright\arraybackslash}p{.12\textwidth}>{\raggedright\arraybackslash}X@{}}
\toprule
\textbf{Method} & \textbf{Constraint class} & \textbf{Exact inference} & \textbf{Learning} & \textbf{CT} & \textbf{Complexity} \\
\midrule
Topology-masked HMM~\citep{rabiner1989tutorial} & local only & yes & EM & yes & no overhead \\
HSMM~\citep{yu2010hidden, hsmm2025recent} & duration only & yes & yes & limited & $O(D_{\max})$ overhead \\
Constrained CRF~\citep{papay2022regular} & regular & yes & conditional\,$^{\dagger}$ & no & polynomial \\
Posterior regularization~\citep{neurips2007posterior} & soft constraints & no (variational) & yes & no & polynomial \\
Probabilistic model checking~\citep{kwiatkowska2022probabilistic} & logic-specific & verification & no\,$^{\ddagger}$ & yes & polynomial in $|\mathcal{C}|$ \\
Weighted transducer~\citep{mohri2002weighted} & regular & yes & model-dependent\,$^{\ddagger}$ & no & polynomial in $|\mathcal{C}|$ \\
Beam search with rejection & cumulative (heuristic) & no (beam) & no & yes & polynomial in beam width \\
Post-hoc filtering~\citep{sridharan2010planning} & cumulative (rejection) & \redit{only if exhaustive} & no & yes & exponential in worst case \\
\midrule
\textbf{CHMM} (this work) & \textbf{regular} & \textbf{yes} & \textbf{yes} & \textbf{jump constraints} & \textbf{polynomial in $|\mathcal{C}|$} \\
\bottomrule
\end{tabularx}
\endgroup
\vspace{0.4em}\\
\begin{minipage}{0.97\textwidth}
\footnotesize
$^{\dagger}$ \redit{Discriminative CRFs optimize conditional likelihood rather than requiring EM. A finite-state acceptor permits cumulative regular constraints in both decoding and training, in which a locally factored score does not limit the support to local constraints.}\\[0.2em]
$^{\ddagger}$ \redit{Weighted automata provide inference operations, whereas learning depends on the probabilistic model and objective attached to their weights. Probabilistic model checking primarily addresses verification.}
\end{minipage}
\end{table*}

\begin{table}[t]
\centering
\caption{Mean per-sequence elapsed decoding time (seconds), over eight sequences at each length in one timing run. Times exclude controller compilation and include neural scoring. The full controller has five memory states, eight reachable augmented states and 18 edges. These measurements describe the implementations used; they are not repeated-seed timing estimates.}
\label{tab:sim-complexity}
\begingroup\fontsize{7.5}{9}\selectfont\setlength{\tabcolsep}{3pt}
\begin{tabular}{lccccc}
\toprule
Method & $T=100$ & $T=200$ & $T=400$ & $T=800$ & HMM ratio \\
\midrule
\textsc{hmm} & $0.00035$ & $0.00073$ & $0.00140$ & $0.00280$ & $1.00\times$ \\
\textsc{tm} & $0.00039$ & $0.00068$ & $0.00138$ & $0.00291$ & $1.04\times$ \\
\textsc{hsmm} & $0.05327$ & $0.21830$ & $0.85087$ & $3.72128$ & $1329.72\times$ \\
\textsc{sp} & $0.00035$ & $0.00064$ & $0.00137$ & $0.00280$ & $1.00\times$ \\
\textsc{phf} & $0.00642$ & $0.01243$ & $0.02547$ & $0.07448$ & $26.61\times$ \\
\textsc{bsr} & $0.03245$ & $0.04562$ & $0.08409$ & $0.28991$ & $103.59\times$ \\
\textsc{bilstm} & $0.00333$ & $0.00630$ & $0.01234$ & $0.02468$ & $8.82\times$ \\
\textsc{bilstm}+\textsc{mask} & $0.01739$ & $0.02761$ & $0.05456$ & $0.11965$ & $42.76\times$ \\
\textsc{bilstm}+\textsc{ctrl} & $0.00242$ & $0.00423$ & $0.00855$ & $0.01772$ & $6.33\times$ \\
\textsc{trf} & $0.00246$ & $0.00588$ & $0.01417$ & $0.04366$ & $15.60\times$ \\
\textsc{trf}+\textsc{mask} & $0.01526$ & $0.02735$ & $0.05584$ & $0.13433$ & $48.00\times$ \\
\textsc{trf}+\textsc{ctrl} & $0.00142$ & $0.00352$ & $0.01117$ & $0.03411$ & $12.19\times$ \\
\textsc{chmm} & $0.00056$ & $0.00098$ & $0.00206$ & $0.00380$ & $1.36\times$ \\
\bottomrule
\end{tabular}
\endgroup
\end{table}

\begin{table*}[t]
\centering
\caption{Parameter recovery against the reference $\theta^{*}$, reported as row-averaged total variation distance (mean across four seeds, with standard deviation in parentheses). The reference is a supervised fit to feasible trajectories, so that these values measure distance to that reference.}
\label{tab:recovery}
\resizebox{\textwidth}{!}{
\begin{tabular}{l c c c c c c}
\toprule
& \multicolumn{2}{c}{$\mathrm{TV}(P, P^{*})$} & \multicolumn{2}{c}{$\mathrm{TV}(B, B^{*})$} & \multicolumn{2}{c}{Iterations} \\
\cmidrule(lr){2-3} \cmidrule(lr){4-5} \cmidrule(lr){6-7}
$N$ & Unconstrained & Constrained & Unconstrained & Constrained & Unconstrained & Constrained \\
\midrule
$10$  & $0.032\,(0.016)$ & $\mathbf{0.007\,(0.003)}$ & $0.083\,(0.003)$ & $\mathbf{0.072\,(0.004)}$ & $13.2$ & $\mathbf{7.8}$ \\
$20$  & $0.029\,(0.007)$ & $\mathbf{0.007\,(0.003)}$ & $0.063\,(0.014)$ & $\mathbf{0.053\,(0.014)}$ & $13.2$ & $\mathbf{7.5}$ \\
$40$  & $0.029\,(0.004)$ & $\mathbf{0.006\,(0.001)}$ & $0.040\,(0.007)$ & $\mathbf{0.036\,(0.005)}$ & $12.8$ & $\mathbf{7.0}$ \\
$80$  & $0.028\,(0.004)$ & $\mathbf{0.004\,(0.001)}$ & $0.033\,(0.004)$ & $\mathbf{0.028\,(0.003)}$ & $12.2$ & $\mathbf{7.0}$ \\
$160$ & $0.027\,(0.003)$ & $\mathbf{0.005\,(0.002)}$ & $0.029\,(0.004)$ & $\mathbf{0.024\,(0.005)}$ & $12.5$ & $\mathbf{7.0}$ \\
\bottomrule
\end{tabular}
}
\end{table*}

\redit{Means and standard deviations for the comparative real-data results are given together in Tables~\ref{tab:drosophila}, \ref{tab:casas} and \ref{tab:har}.}

\section{Figures}
This section provides formal figures addressed in the main text. 

\begin{figure}[t]
\centering
\includegraphics[width=\columnwidth]{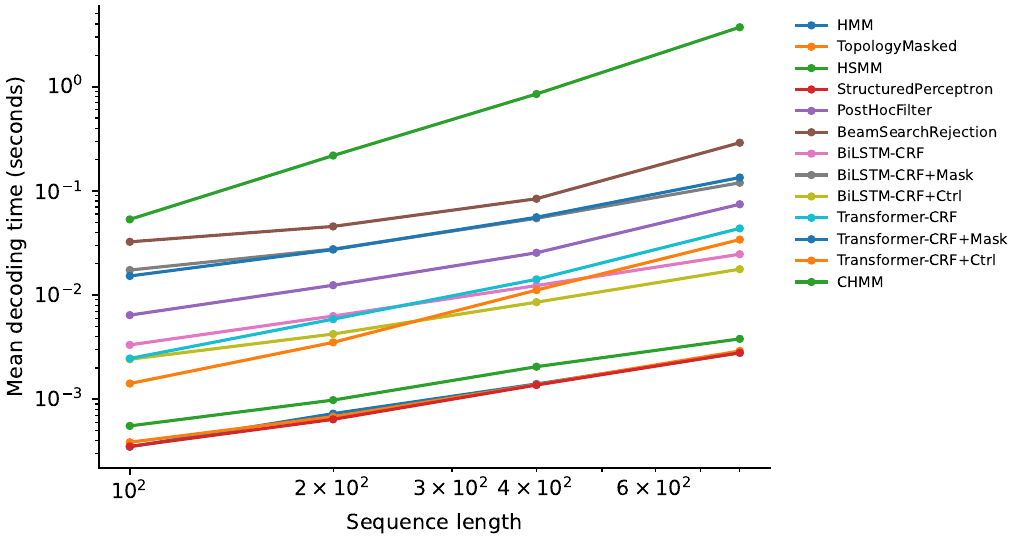}
\caption{\redit{Measured decoding times for 13 implementations, averaged over 8 sequences at each length in one timing run. The reachable controller size is fixed. Neural times include scoring, controller compilation is excluded. The HSMM exhibits substantially steeper scaling than the HMM and CHMM on these lengths.}}
\label{fig:sim-scaling}
\end{figure}


\clearpage
\bibliography{refs}

\end{document}